\documentclass{aa}  

\usepackage{graphicx,aas_macros}
\usepackage{txfonts}
\usepackage[colorlinks=true,allcolors=blue]{hyperref}
\usepackage{xcolor}
\usepackage{textgreek}
\graphicspath{Immagini/}

\begin{document} 

\newcommand{\solarM}{\,\mathrm{M}_\odot}
\newcommand{\solarL}{\,\mathrm{L}_\odot}
\newcommand{\E}[1]{\times10^{#1}}
\newcommand{\nH}{\,\mathrm{cm}^{-2}}
\newcommand{\ergcms}{\,\mathrm{erg}\,\mathrm{cm}^{-2}\,\mathrm{s}^{-1}}
\newcommand{\asec}{\,\mathrm{arcsec}}
\newcommand{\amin}{\,\mathrm{arcmin}}
\newcommand{\magnitude}{\,\mathrm{mag}}
\newcommand{\ergs}{\,\mathrm{erg}\,\mathrm{s}^{-1}}
\newcommand{\pdotunit}{\,\mathrm{s}\,\mathrm{s}^{-1}}
\newcommand{\LX}{L_\mathrm{X}}
\newcommand{\LEdd}{L_\mathrm{Edd}}
\newcommand{\pspin}{P_\mathrm{spin}}
\newcommand{\porb}{P_\mathrm{orb}}
\newcommand{\asini}{a_\mathrm{x}\sin i}
\newcommand{\FX}{F_\mathrm{X}}
\newcommand{\apm}[2]{_{-#1}^{+#2}}
\newcommand{\xmm}{\emph{XMM-Newton}}
\newcommand{\swift}{\emph{Swift}}
\newcommand{\chandra}{\emph{Chandra}}
\newcommand{\erosita}{\emph{eROSITA}}
\newcommand{\nustar}{\emph{NuSTAR}}
\newcommand{\ulxie}{ULX-7}
\newcommand{\salt}{\emph{SALT}}
\newcommand{\rxte}{\emph{RXTE}}
\newcommand{\ulxielong}{M51\,ULX-7}
\newcommand{\obsidst}{0883550101}
\newcommand{\obsidnd}{0883550201}
\newcommand{\obsidrd}{0883550301}
\newcommand{\obsidold}{0303420201}
\newcommand{\matt}[1]{\textcolor{red}{#1 - MI}}
\newcommand{\sem}[1]{\textcolor{magenta}{#1 - SEM}}
\newcommand{\giallo}[1]{\textcolor{orange}{#1 - GLI }}
\newcommand{\paolo}[1]{\textcolor{green}{#1 - PE}}
\newcommand{\rob}[1]{\textcolor{cyan}{#1 - RA}}
\newcommand{\rev}[1]{\textbf{#1}}

   \title{Skipping a beat: discovery of persistent quasi-periodic oscillations associated with pulsed fraction drop of the spin signal in M51\,ULX-7}
   \titlerunning{M51\,ULX-7 variability}
   \authorrunning{M. Imbrogno et al.}

   \author{Matteo Imbrogno
          \inst{1,2,3}
          \and
          Sara Elisa Motta
          \inst{4,5}
          \and
          Roberta Amato
          \inst{2}
          \and
          Gian Luca Israel
          \inst{2}
          \and
          Guillermo Andres Rodr\'{i}guez Castillo
          \inst{6}
          \and
          Murray Brightman
          \inst{7}
          \and
          Piergiorgio Casella
          \inst{2}
          \and
          Matteo Bachetti
          \inst{8}
          \and
          Felix F{\"u}rst
          \inst{9}
          \and
          Luigi Stella
          \inst{2}
          \and 
          Ciro Pinto
          \inst{6}
          \and
          Fabio Pintore
          \inst{6}
          \and
          Francesco Tombesi
          \inst{1,2,10}
          \and
          Andr{\'e}s G{\'u}rpide
          \inst{11}
          \and
          Matthew J. Middleton
          \inst{11}
          \and
          Chiara Salvaggio
          \inst{4}
          \and
          Andrea Tiengo
          \inst{12,13}
          \and
          Andrea Belfiore
          \inst{13}
          \and
          Andrea De Luca
          \inst{13}
          \and
          Paolo Esposito
          \inst{12,13}
          \and
          Anna Wolter
          \inst{14}
          \and
          Hannah P. Earnshaw
          \inst{7}
          \and
          Dominic J. Walton
          \inst{15}
          \and
          Timothy P. Roberts
          \inst{16}
          \and
          Luca Zampieri
          \inst{17}
          \and
          Martino Marelli
          \inst{13}
          \and
          Ruben Salvaterra
          \inst{13}
          }

   \institute{Dipartimento di Fisica, Università degli Studi di Roma “Tor Vergata”, via della Ricerca Scientifica 1, I-00133 Rome, Italy\\ 
              \email{matteo.imbrogno@inaf.it}
         \and 
             INAF -- Osservatorio Astronomico di Roma, via Frascati 33, I-00078 Monte Porzio Catone (RM), Italy
         \and 
            Dipartimento di Fisica, Università degli Studi di Roma “La Sapienza”, piazzale Aldo Moro 5, I-00185 Roma, Italy
         \and 
             INAF -- Osservatorio Astronomico di Brera, via E. Bianchi 46, I-23807 Merate (LC), Italy
         \and 
             Astrophysics Sub-department, Department of Physics, University of Oxford, Denys Wilkinson Building, Keble Road, Oxford OX1 3RH, UK
         \and 
            INAF/IASF Palermo, via Ugo La Malfa 153, I-90146 Palermo, Italy
         \and 
             Cahill Center for Astrophysics, California Institute of Technology, 1216 East California Boulevard, Pasadena, CA 91125, USA
        \and 
            INAF -- Osservatorio Astronomico di Cagliari, via della Scienza 5, I-09047 Selargius (CA), Italy
        \and 
            European Space Astronomy Centre (ESAC), ESA, Camino Bajo del Castillo s/n, Villanueva de la Ca\~{n}ada, E-28692 Madrid, Spain
        \and 
            INFN – Roma Tor Vergata, via della Ricerca Scientifica 1, I-00133 Rome, Italy
        \and 
            Department of Physics and Astronomy, University of Southampton, Highﬁeld, Southampton SO17 1BJ, UK
        \and 
            Scuola Universitaria Superiore IUSS Pavia, Palazzo del Broletto, piazza della Vittoria 15, I-27100 Pavia, Italy
        \and 
            INAF, Istituto di Astrofisica Spaziale e Fisica Cosmica, via Alfonso Corti 12, I-20133, Milano, Italy
        \and 
            INAF -- Osservatorio Astronomico di Brera, via Brera 28, I-20121 Milano, Italy
        \and 
        Centre for Astrophysics Research, University of Hertfordshire, College Lane, Hatfield AL10 9AB, UK
        \and 
        Centre for Extragalactic Astronomy \& Dept of Physics, Durham University, South Road, Durham DH1 3LE, UK
        \and 
        INAF – Osservatorio Astronomico di Padova, Vicolo dell’Osservatorio 5, I-35122 Padova, Italy
             }

   \date{Received MONTH XX, YYYY; accepted MONTH XX, YYYY}

 
  \abstract
   { 
   The discovery of pulsations in (at least) six ultraluminous X-ray sources (ULXs) has shown that neutron stars can accrete at (highly) super-Eddington rates, challenging the standard accretion theories. \ulxielong, with a spin signal of $P\simeq2.8$\,s, is the pulsating ULX (PULX) with the shortest known orbital period ($P_\mathrm{orb}\simeq2$\,d) and has been observed multiple times by \xmm, \chandra\ and \nustar.
   }
   { 
   We report on the timing and spectral analyses of three \xmm\ observations of \ulxielong\ performed between the end of 2021 and the beginning of 2022, together with a timing re-analysis of \xmm, \chandra\ and \nustar\ archival observations. 
   }
   {
   We investigated the spin signal by applying accelerated search techniques and studied the power spectrum through the Fast Fourier Transform, looking for (a)periodic variability in the source flux. We analysed the energy spectra of the 2021--2022 observations and compared them to the older ones.
   }
   {
   We report the discovery of a recurrent, significant ($>$3$\sigma$) broad complex at mHz frequencies in the power spectra of \ulxielong. We did not detect the spin signal, setting a 3$\sigma$ upper limit on the pulsed fraction of $\lesssim10\%$ for the single observation. The complex is significantly detected also in five \chandra\ observations performed in 2012.  
  }
   {
   \ulxielong\ represents the second PULX for which we have a significant detection of mHz-QPOs at super-Eddington luminosities. These findings suggest that one should avoid using the observed QPO frequency to infer the mass of the accretor in a ULX. The absence of spin pulsations when the broad complex is detected suggests that the mechanism responsible for the aperiodic modulation also dampens the spin signal's pulsed fraction. If true, this represents an additional obstacle in the detection of new PULXs, suggesting an even larger occurrence of PULXs among ULXs.
   }

   \maketitle
%

\section{Introduction}

First detected by the \textit{Einstein} mission at the end of the 1970s in nearby galaxies \citep{Fabbiano1989}, ultraluminous X-ray sources (ULXs) are off-nuclear, point-like, accreting objects whose X-ray luminosity (under the assumption of isotropic emission) is in excess of 10$^{39}\ergs$ \citep[see][for recent reviews]{Kaaret2017,Fabrika2021,King2023,Pinto2023}, i.e. the Eddington luminosity ($\LEdd\simeq1.3\E{38}M/M_\odot\ergs$, where $M$ is the accretor mass) of a $\sim$10\,M$_\odot$ black hole (BH). Historically, their extreme X-ray luminosity (which can be as high as 10$^{42}\ergs$) was explained within the context of sub-Eddington accretion onto intermediate-mass BHs \citep[IMBHs, with $M_\mathrm{BH}\simeq10^2-10^6\solarM$; see e.g.][]{Colbert1999}. In this scenario, ULXs would be scaled-up versions of the Galactic BH binaries (GBHBs). For these systems, the presence of quasi-periodic oscillations (QPOs) at 0.1--15\,Hz (type-C QPOs) in the power density spectrum (PDS) has been proposed to provide a tool for an indirect estimate of the BH mass through timing analysis \citep{Casella2005,Casella2008}. QPOs in the mHz range have been detected in the PDSs of various ULXs: under the assumption that these QPOs are the low-frequency equivalent of type-C QPOs seen in GBHBs, the derived masses are consistent with those expected from IMBHs \citep[see e.g.][]{Strohmayer2007,Strohmayer2009,Pasham2015}. 

Back in the early 2000s, it was proposed that ULXs could represent a class of stellar-mass, super-Eddington accretors, both NSs and BHs \citep{King2001,Poutanen2007,Zampieri2009}. This possibility received a first confirmation when \cite{Bachetti2014} detected coherent pulsations with period $\pspin\simeq1.37$\,s in the X-ray flux of M82 X-2, clearly identifying this source as an accreting, spinning, magnetic neutron star (NS) with $\LX\gtrsim10^{39}\ergs$, i.e. a pulsating ULX (PULX). 
The subsequent discovery of other extragalactic PULXs \citep{Furst2016,Israel2017a,Israel2017,Carpano2018,Sathyaprakash2019,RodriguezCastillo2020} has shown that accreting compact objects can exceed $\LEdd$ by up to a factor 500. 

Only 6 extragalactic PULXs mostly emitting at super-Eddington luminosities are currently known out of more than 1800 (confirmed and candidates) ULXs \citep{Walton2022,Tranin2024}, but the hypothesis that the population of NS-powered ULXs is higher is supported by both spectral and timing results (as well as other 6 known NSs which exhibited $\ge10^{39}$ erg\,s$^{-1}$ outbursts for short periods, see \citealt{King2023} and references therein). \cite{RodriguezCastillo2020} noted that, if one considers only the ULXs for which we have enough statistics to detect spin pulsations with similar properties, the ratio of (confirmed and unknown) PULXs to the whole (confirmed and candidates) ULX population is $\gtrsim$25\%. Additionally, \cite{Koliopanos2017}, \cite{Pintore2017} and \cite{Walton2018} found that PULX X-ray spectra are practically indistinguishable from the X-ray spectra of many ULXs, suggesting a larger population of ULXs powered by accreting NSs. The super-Eddington nature of these systems is supported by the magnitude of the decay of the orbit of M82 X-2 \citep{Bachetti2022} and by the magnitude of the spin-up rate of NGC 5907 ULX-1 \citep{Israel2017}. 
The discovery of pulsations from M82 X-2 has also shown that, when interpreting the mHz-QPOs of ULXs as proxy of the accretor mass, the association with type-C QPOs is not always valid: \cite{Feng2010} found a QPO in the PDS of the PULX M82 X-2 (source X42.3+59 in their article) at a frequency $\nu_\mathrm{QPO}\simeq3$--4\,mHz, which led them to identify M82 X-2 as an IMBH with a mass of $\sim$12,000--43,000\,M$_\odot$. 

\ulxielong\ (\ulxie\ hereafter) has been identified as a PULX when \cite{RodriguezCastillo2020} discovered coherent pulsations at a period $\pspin\simeq2.8$\,s and a spin-up rate $\dot{P}_\mathrm{spin}\simeq-2.4\E{-10}$\,s\,s$^{-1}$. It is also one of the four PULXs, together with NGC 7793 P13 \citep{Furst2018,Furst2021}, M82 X-2 \citep{Bachetti2022} and NGC 5907 ULX-1 \citep{Israel2017a,Belfiore2024}, with a confirmed orbital solution ($\porb\simeq2$\,d, projected semi-major axis $\asini\simeq28$\,lt-s, \citealt{RodriguezCastillo2020}), and with a massive ($>8\solarM$) companion star. The source flux shows periodic dips, associated with the orbital periods, suggesting a system inclination angle of $i\sim60^\circ$ \citep{Hu2021,Vasilopoulos2021}. The X-ray flux shows a super-orbital modulation, a common property among (P)ULXs \citep[see e.g.][]{Lin2015,Furst2018,Weng2018}. Initially detected by \swift\ at a period of $\simeq38$\,d \citep{Brightman2020}, it has since evolved towards a 44\,d-long period \citep{Brightman2022}. This evolution, together with an inclination of the disk components dependent on the super-orbital phase, supports the scenario of a precessing disk. 

In this paper, we report the detection of a QPO-like modulation in the X-ray flux of \ulxie\ in three different \xmm\ observations performed in 2021/2022. These findings represent the first unambiguous detection of QPOs in this source (with tens of cycles sampled)  over a baseline of about one month. 
\ulxie\ is therefore the second PULX for which we have the detection of a QPO at super-Eddington luminosities.

The article is structured as follows: in Sect.\,\ref{sec:ObsDataReduction} we describe the observations analysed in this article and the data processing techniques that we applied. In Sect.\,\ref{sec:results}, we report on the results of our timing and spectral analyses. We discuss our results and the possible nature of this quasi-periodic modulation in Sect.\,\ref{sec:Discussion}, while we draw our conclusions in Sect.\,\ref{sec:Conclusioni}.

\section{Observations and Data Reduction}\label{sec:ObsDataReduction}

The M51 field has been observed 45 times with relatively deep, pointed X-ray observations, including 14 observations from \xmm\ \citep{Jansen2001}, 27 from \chandra\ \citep{Weisskopf2000} and 4 from \nustar\ \citep{Harrison2013}. For our analysis, we excluded those observations during which \ulxie\ was not detected or detected with less than 100 counts. We also excluded observations lasting less than 20\,ks, to ensure the detection of a significant number of cycles of the modulation highlighted above, and observations whose PDSs were dominated by \chandra\ dithering, which introduces spurious signals at $P_\mathrm{spurious}\simeq700-1000$\,s\footnote{\url{https://cxc.cfa.harvard.edu/ciao/why/dither.html}} \citep[see e.g. Sec.~2.3 of][]{Nichols2010}. 
In Table~\ref{tab:XrayObs} we list the remaining 27 observations (12 from \xmm, 14 from \chandra\ and 1 from \nustar) we analysed and discuss in this work.
We used the \chandra\ position ($\mathrm{R.A.}=13^\mathrm{h}30^\mathrm{m}01\fs02$, $\mathrm{Dec}=47^\circ13'43\farcs8$, J2000; \citealt{Kuntz2016}) to convert the event arrival times to the barycentre of the Solar System and extract source events for both \xmm\ and \chandra. Unless otherwise stated, in this work the reported errors correspond to 1$\sigma$ (68.3\%) confidence ranges.

\begin{table}
    \centering
    \caption{Log of \xmm\ and \chandra\ observations of \ulxielong\ analysed in this work.}
    \resizebox{\columnwidth}{!}{
    \begin{tabular}{cccc}
    \hline
    \hline
     Satellite & ObsID & Start Date & Exposure\tablefootmark{a}\\ 
      &  &  & (ks)\\ 
    \hline
        \xmm & 0112840201 & 2003 Jan 15 & 20.9 \\
        \chandra & 3932 & 2003 Aug 07 & 48.0 \\
        \xmm & 0212480801 & 2005 Jul 1 & 49.2 \\
        \xmm & 0303420101 & 2006 May 20 & 54.1 \\
        \xmm & 0303420201 & 2006 May 24 & 36.8 \\
        \chandra\tablefootmark{\dag} & 13813 & 2012 Sep 9 & 179.2 \\
        \chandra\tablefootmark{\dag} & 13812 & 2012 Sep 12 & 157.5 \\
        \chandra\tablefootmark{\dag} & 15496 & 2012 Sep 19 & 41.0 \\
        \chandra\tablefootmark{\dag} & 13814 & 2012 Sep 20 & 189.9 \\
        \chandra\tablefootmark{\dag} & 13815 & 2012 Sep 23 & 67.2 \\
        \chandra & 13816 & 2012 Sep 26 & 73.1 \\
        \xmm & 0824450901 & 2018 May 13 & 78.0 \\
        \xmm & 0830191401 & 2018 May 25 & 98.0 \\
        \xmm & 0830191501 & 2018 Jun 13 & 63.0 \\
        \xmm & 0830191601 & 2018 Jun 15 & 63.0 \\
        \nustar & 60501023002 & 2019 Jul 10 & 162.0 \\
        \xmm & 0852030101 & 2019 Jul 12 & 77.0 \\
        \chandra & 23472 & 2020 Oct 13 & 33.6 \\
        \chandra & 23474 & 2020 Dec 21 & 36.1 \\
        \chandra & 23475 & 2021 Jan 28 & 34.5 \\
        \chandra & 23476 & 2021 Mar 1 & 34.4 \\
        \chandra & 23479 & 2021 Jun 7 & 35.0 \\
        \chandra & 23480 & 2021 Jul 13 & 34.5 \\
        \xmm\tablefootmark{\dag} & \obsidst\tablefootmark{b} & 2021 Nov 22 & 130.4 \\
        \xmm\tablefootmark{\dag} & \obsidnd\tablefootmark{b} & 2021 Nov 24 & 130.2 \\
        \xmm\tablefootmark{\dag} & \obsidrd\tablefootmark{b} & 2022 Jan 7 & 131.4 \\
    \hline
    \hline
    \end{tabular}
    }
    \tablefoot{
    \tablefoottext{a}{Pre-flare filtering exposure time.}
    \tablefoottext{b}{Observations analysed for the first time for this paper.}
    \tablefoottext{\dag}{Observations that show the ks-long aperiodic modulation in the power density spectrum with a significance $\geq3\sigma$.}
    }
    \label{tab:XrayObs}
\end{table}

    \subsection{XMM-Newton}\label{sec:XMMreduction}

    As part of the \xmm\ Large Program "Too B or not too B" (LP hereafter) we observed the field of view of M51, with \ulxie\ on-axis, three times between November 2021 and January 2022, for a total (nominal) exposure time of about 390\,ks. The \xmm\ observations with ObsID \obsidst, \obsidnd\ and \obsidrd\ have not yet been presented elsewhere. Hereafter, we refer to observations \obsidst, \obsidnd\ and \obsidrd\ as observations A, B, and C, respectively. \ulxie\ was detected in every observation in the three CCD cameras, with the EPIC PN \citep{Struder2001} operated in Full Frame mode (time resolution $\delta t=73.4$\,ms) and both EPIC MOS \citep{Turner2001} in Small Window mode (time resolution $\delta t=0.3$\,s for the central CCD) to resolve the 2.8\,s-long spin pulsations. We used \textsc{SAS} \citep{Gabriel2004} v21.0.0 with the latest \xmm\ calibrations and applied standard data reduction procedures to prepare the raw data for both timing and spectral analysis. We selected only the events with $\texttt{PATTERN}\leq4$ from the EPIC PN data and events with $\texttt{PATTERN}\leq12$ from the EPIC MOS data. We considered events in the 0.3--10\,keV band for both our timing and spectral analysis. Considering the presence of nearby sources, we selected events for both timing and spectral analysis from a circular region with 20\arcsec-radius centred on the source position. To properly take into account the diffuse emission in the proximity of \ulxie\ and a close chip gap, the background was estimated from an annular region centred on the source position and with inner and outer radii equal to 21\arcsec and 39\arcsec, respectively.
There are X-ray sources within this region, which were excluded from the event selection. Such a small inner radius is usually not recommended for the background region, due to the \xmm\ point spread function. We verified that the contamination from the source is acceptable by considering other circular background regions further away from \ulxie. We found that both the background-subtracted light curve count rates and the spectra parameters are consistent with the ones produced using the annular region. Since the latter better describes the diffuse emission surrounding \ulxie, we preferred our initial choice for the annular background region.
    
    We extracted the high energy ($E > 10$\,keV) light curves of the entire field of view to verify the presence of high-background particle flares. To correct for the high-flaring background our observations were affected by, we adopted two different criteria for the timing and spectral analysis. In the first case, considering that the background does not affect the search for the spin signal and to avoid introducing too many gaps in the light curve, we removed only the particle flares at the beginning and/or at the end of the observation. The effective exposure time of our three observations is then reduced to 117.0, 114.1, and 127.1\,ks, respectively. The event arrival times were barycentred to the source coordinates using the \textsc{SAS} task \texttt{barycen}. The data for timing analysis were background-corrected by the means of the \textsc{SAS} task \texttt{epiclccorr}. For the spectral analysis of observations \obsidnd\ and \obsidrd, we further removed high background intervals occurring in the midst of the observations. As a consequence, the net exposure time for the spectral analysis is further reduced to 72.0, 101.5 and 102.6\,ks during observation \obsidnd\ for the EPIC PN, EPIC MOS1 and EPIC MOS2 data, respectively. In observation \obsidrd, the net exposure is  93.3, 121.3 and 122.9 ks for the EPIC PN, EPIC MOS1 and EPIC MOS2 data, respectively. 
    We discarded the spectra of observation \obsidst\ due to a particularly high particle flare contamination. We created response matrices and ancillary files with the \textsc{SAS} tasks \texttt{rmfgen} and \texttt{arfgen}. Spectra were binned with a minimum of 25 counts per energy bin to allow for fits with a $\chi^2$ statistics.

    We followed the same data reduction procedure (PATTERN selection, background filtering, barycentric correction and energy spectra extraction) for the archival \xmm\ observations reported in Table~\ref{tab:XrayObs}.

    \subsection{Chandra}\label{sec:Chandrareduction}

    We downloaded the latest version of the archival data of the M51 field observations. For the data reduction of \chandra\ observations, we employed the \chandra\ Interactive Analysis of Observations (\textsc{ciao}) software v4.15 \citep{Fruscione2006} and v4.10.7 of the calibration database. We reprocessed the data with the task \texttt{chandra\_repro} and we applied the barycentric correction with \texttt{axbary}. We then used the \texttt{wavdetect} script to verify that \ulxie\ had been detected during the observation and to estimate the source extraction region. 
    We extracted the source events by the means of the task \texttt{dmcopy}. We considered events in the 0.5--10\,keV band for our timing analysis. The radius of the source extraction region depends on the observation, but we verified that it was always smaller than 5\arcsec. Therefore, we evaluated the background using an annular region centred on \ulxie\ and with inner (outer) radius 5\arcsec (20\arcsec) for all the observations. 

    \subsection{NuSTAR}\label{sec:NuSTARreduction}

    For the data reduction of the \nustar\ observation of \ulxie, we followed \cite{Brightman2022}. We used \texttt{nupipeline} with \texttt{saacalc=3 saamode=OPTIMIZED tentacle=yes} to produce cleaned and calibrated events of observation 60501023002, resulting in a net exposure time of 162.0\,ks. We selected these settings to account for enhanced background during passages of the South Atlantic Anomaly. We extracted source events in a circular region with a radius of 30\arcsec, while background events were extracted from a circular region on the same chip with a radius of 100\arcsec. We applied the barycentric correction to the event times of arrival using the \texttt{barycorr} task.


\section{Data Analysis and Results}\label{sec:results} 

    \subsection{Timing Analysis}\label{sec:LPTiming}

\begin{figure*}
        \centering
        \resizebox{\hsize}{!}
        {\includegraphics{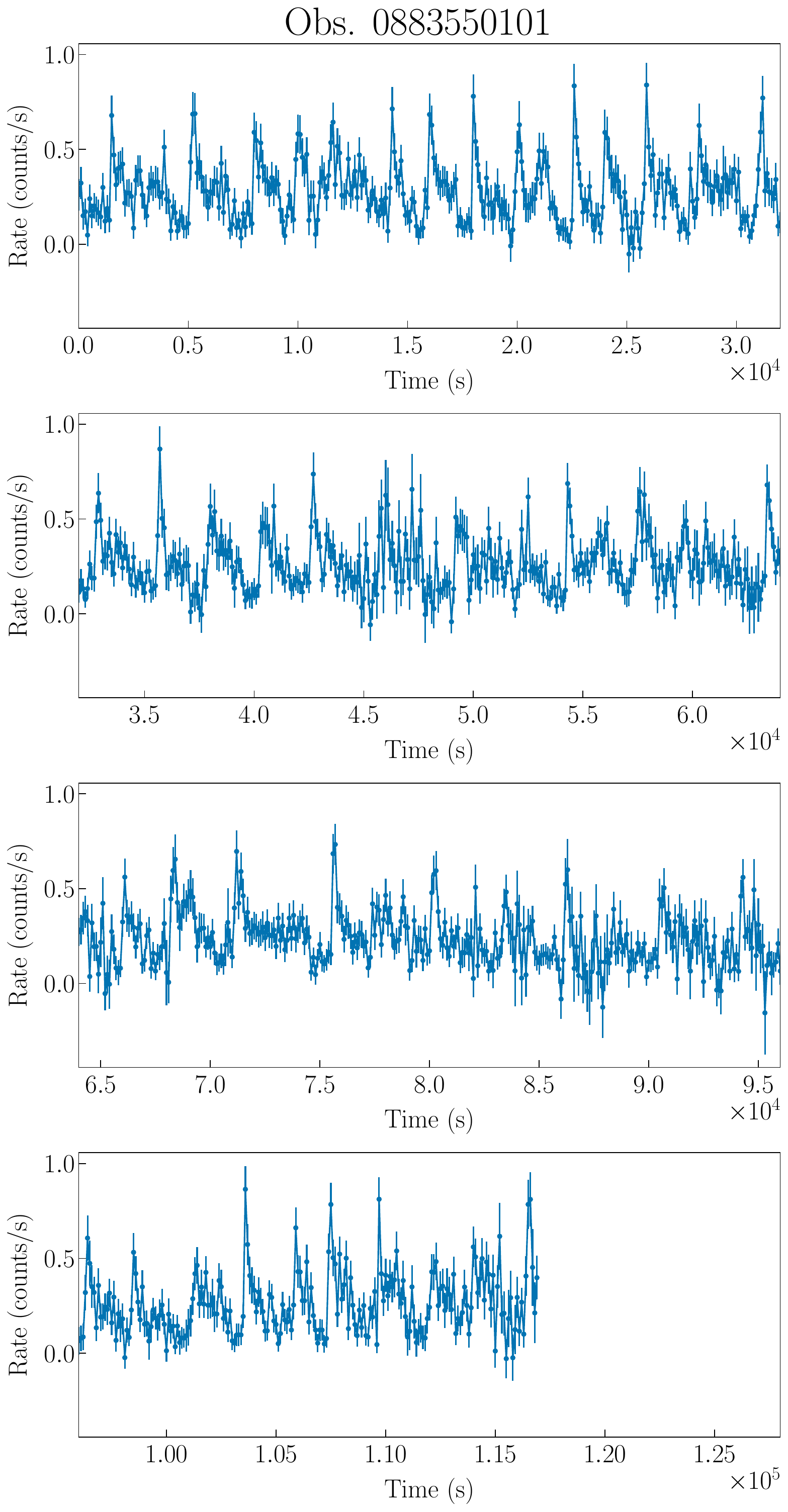}\qquad
        \includegraphics{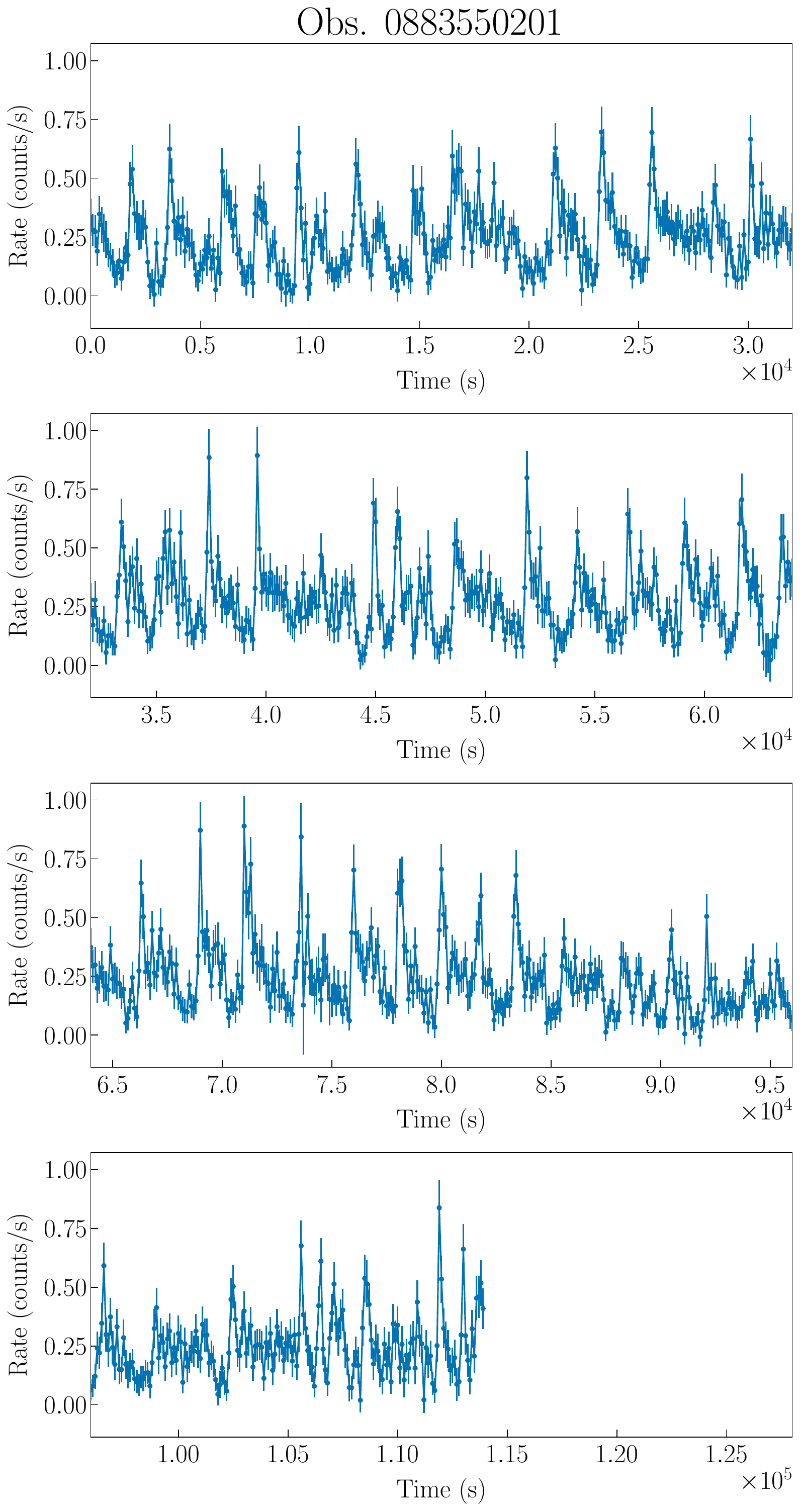}\qquad
        \includegraphics{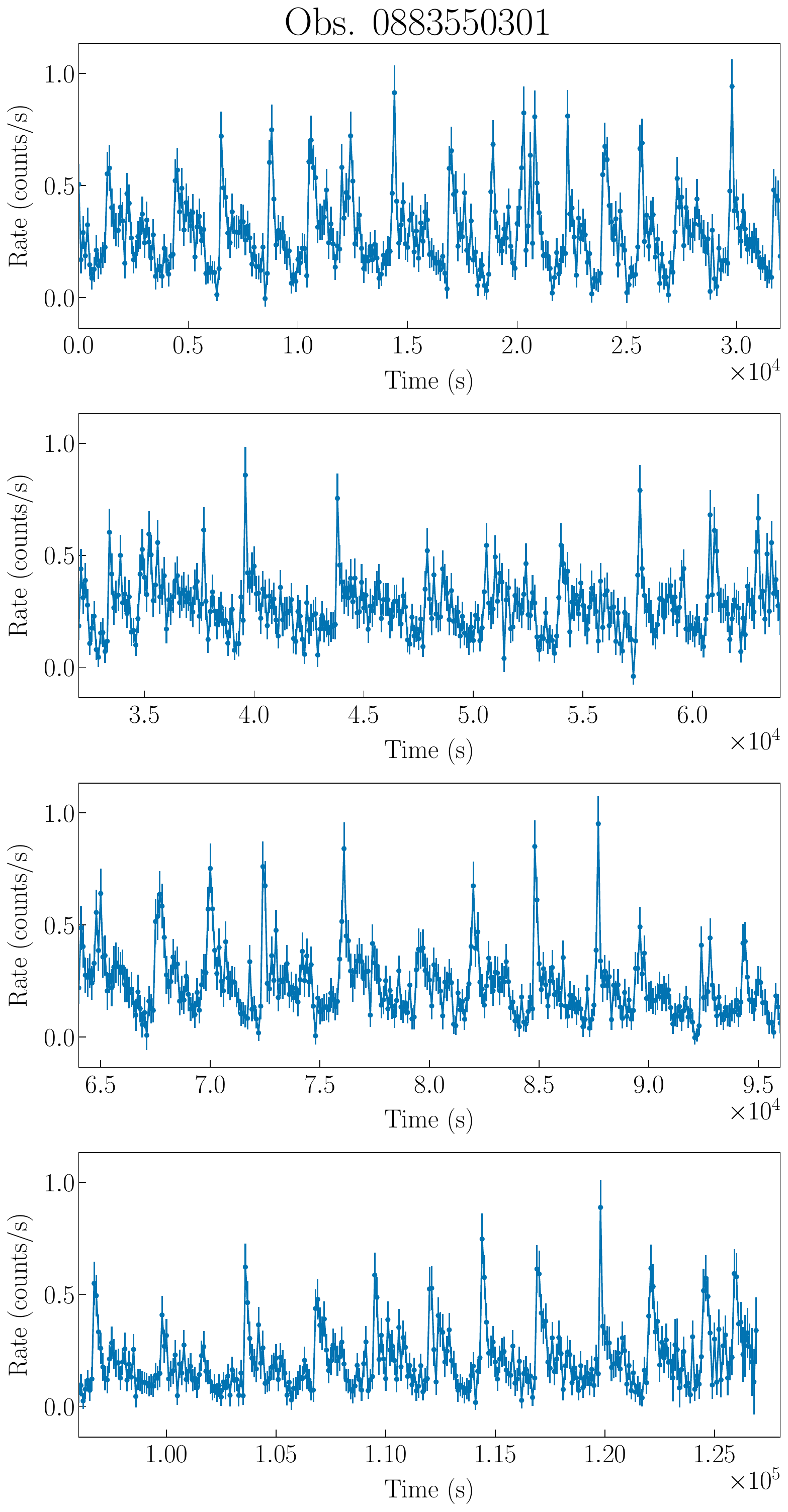}}
        \caption{PN+MOS light curve of \ulxie\ in the 0.3--10\,keV band in the three observations of our LP (bin time of 100\,s). Each column corresponds to a different observation, while each row shows 32\,ks-long chunks of the corresponding observation. From left to right: observations A, B, C. 
        }
        \label{fig:lc_all_obs}
    \end{figure*}

    We first produced the 0.3--10\,keV PN+MOS light curves for each \xmm\ LP observation, with a bin time of 100\,s. 
    We report the three light curves in Fig.~\ref{fig:lc_all_obs}, where each column corresponds to a different observation. A recurring, ks-long flaring feature is clearly visible in all three light curves. To study its evolution in energy we considered a soft 0.3--1.5\,keV band and a hard 1.5--10\,keV band 
    and produced the corresponding light curves following the same procedure. We chose to divide the two bands at 1.5\,keV for several reasons. First of all, with this choice each band includes approximately half of the detected events. Secondly, as one can see from Fig.~\ref{fig:FitSpectraResiduals}, it is also approximately the energy at which the curvature of the energy spectra changes, suggesting that the hard component is starting to dominate the emission. The soft (hard) band, therefore, allows us to probe the outer (inner) regions of the disk. 
   
    We computed the hardness ratio $HR=C_\mathrm{h}/C_\mathrm{s}$, where $C_\mathrm{h}$ ($C_\mathrm{s}$) is the count rate in the hard (soft) band. The time evolution of the hardness ratio is consistent with it being constant in time. Other choices for the soft/hard band (0.3--1/1--10\,keV, 0.3--2/2--10\,keV) show the same trend, suggesting that the available data do not allow us to identify a spectral evolution. 

\begin{figure*}
         \centering
            \includegraphics[scale=0.21]{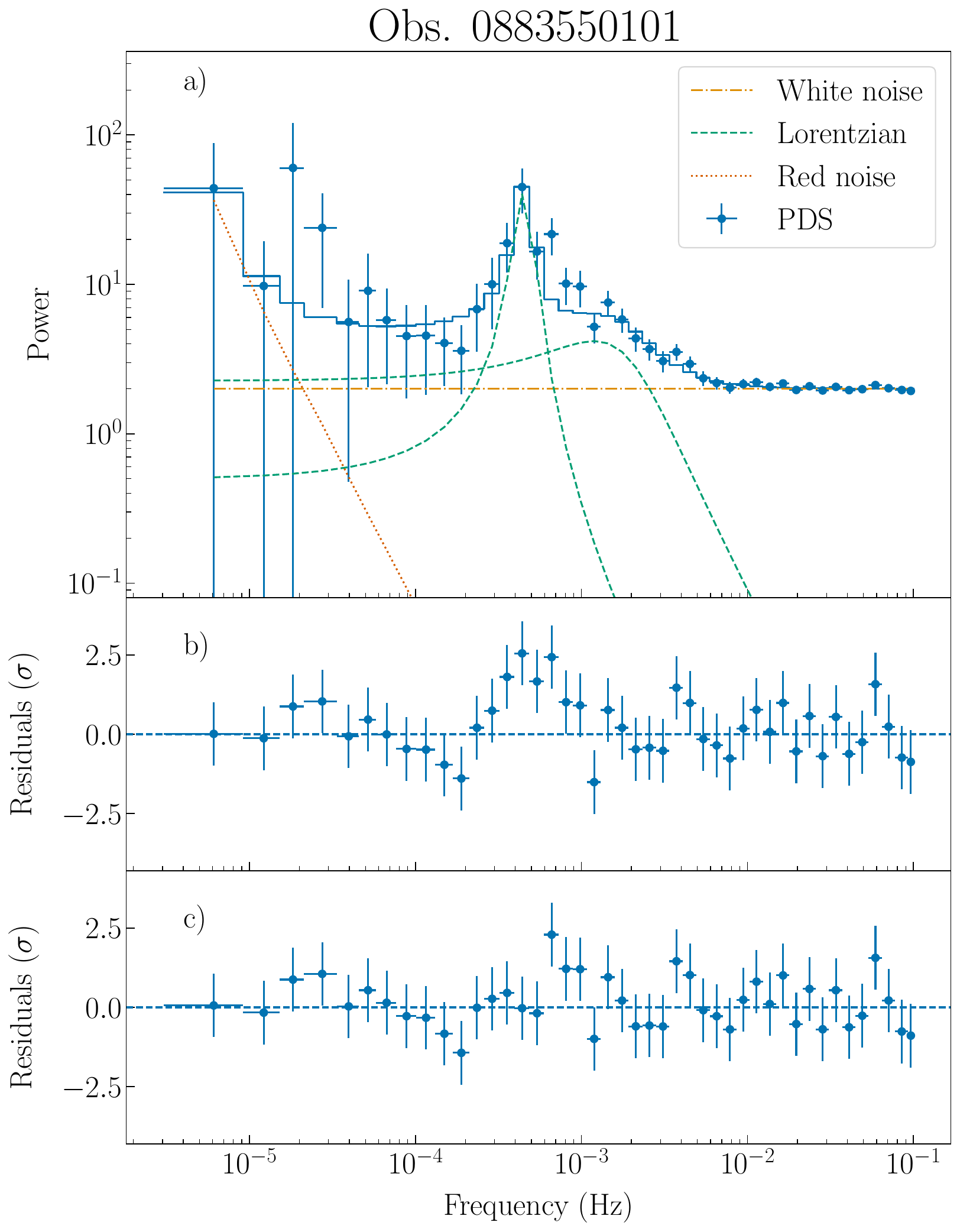}\quad\includegraphics[scale=0.21]{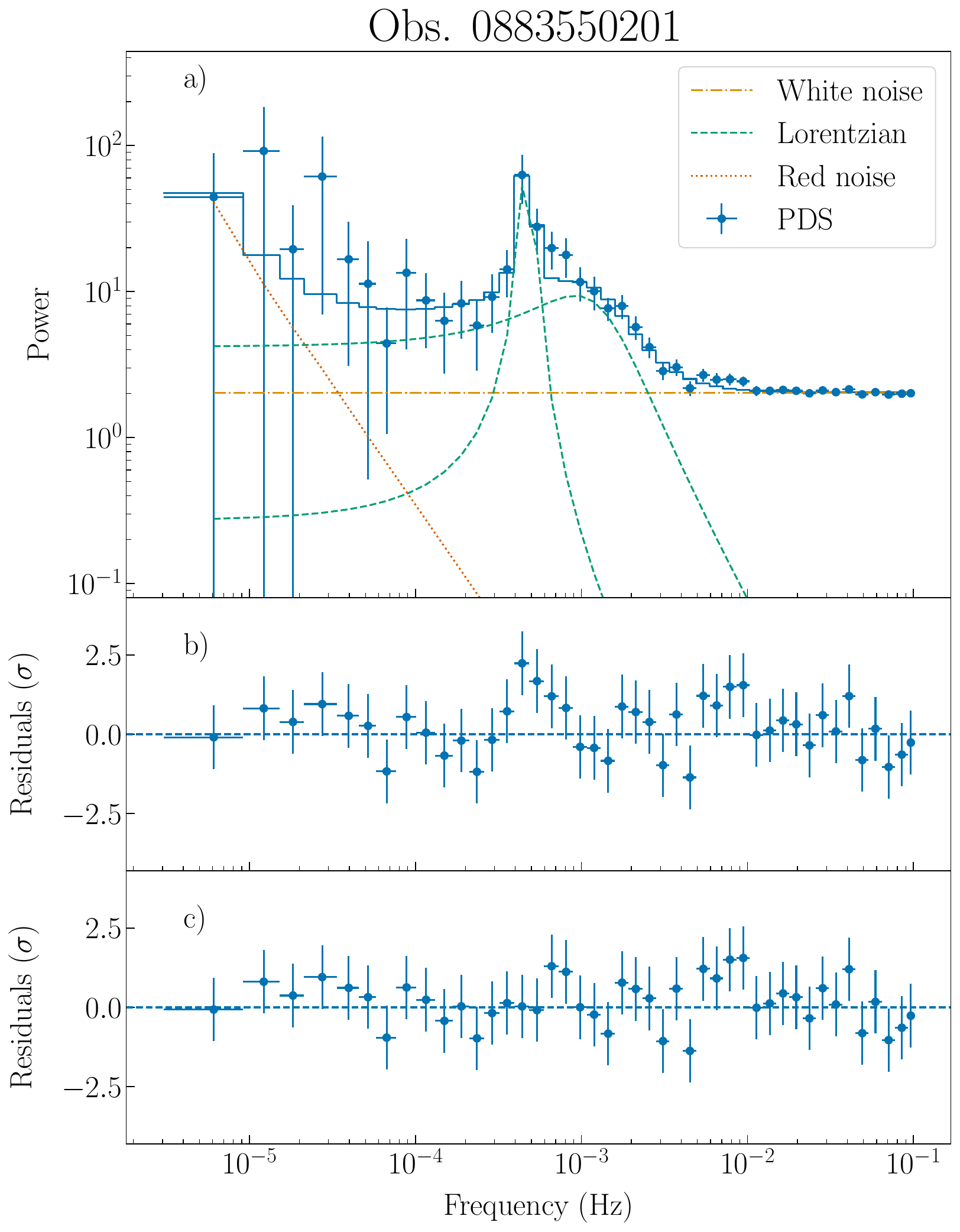}\quad\includegraphics[scale=0.21]{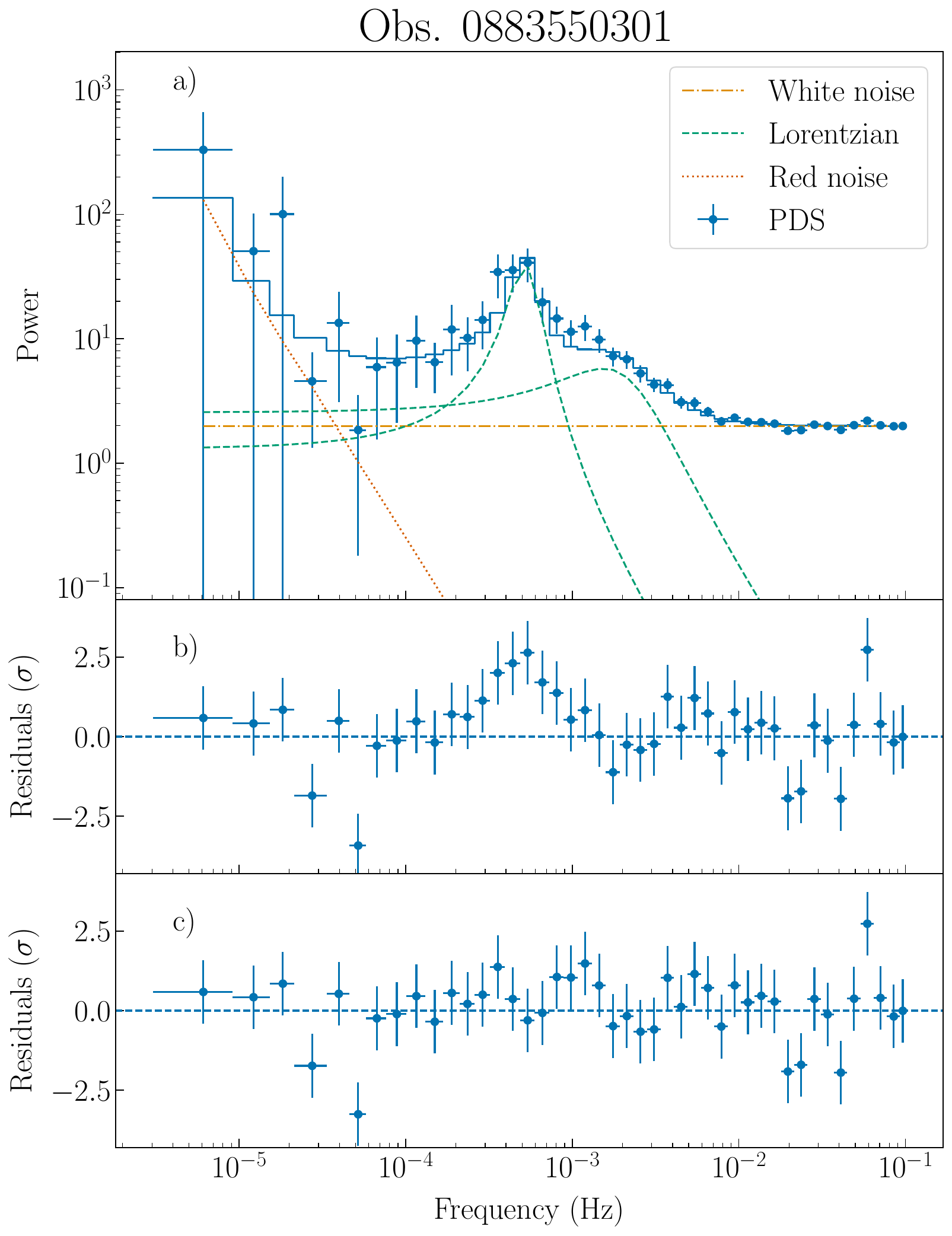}
       \caption{PDSs in the 0.3--10\,keV band of the LP observations A (left), B (centre), C (right). Panel a: PDS with the components used for the fit. Panel b: residuals of the model without the second Lorentzian. Panel c: residuals of the final model with two Lorentzians.  For each observation, we combined data from both PN and MOS cameras. The PDSs are computed using the Leahy normalization. The model we used to fit the PDSs is described by equation~\eqref{eq:PDSmodel}.
               }
          \label{fig:psds}
    \end{figure*}

    In Fig.~\ref{fig:psds} we report the 0.3--10\,keV power density spectra (PDSs) corresponding to each of the three LP observations, generated with the \texttt{powspec} task in the \textsc{XRONOS} package \citep{Stella1992}, included in the \textsc{HEASoft} v6.32.1\footnote{\url{https://heasarc.gsfc.nasa.gov/docs/software/lheasoft/}}. The PDSs were computed with a bin time of 5\,s for two reasons: 1) to have a range of frequencies where we could constrain the white noise component of the fit and 2) to avoid cutting the right shoulder of the broad feature, which extends up to $\sim$0.01\,Hz. For these PDSs, we adopted the Leahy normalization \citep{Leahy1983} and a logarithmic rebin factor of 1.20 (i.e. each bin is 20\% larger than the previous one). A broad feature in the (sub-)mHz range, likely associated with the flare-like activity in the light curves, is present in each PDS. We converted the PDSs in \textsc{XSPEC} format to fit them with \textsc{XSPEC} \citep{Arnaud1996} v12.13.1, included in the same \textsc{HEASoft} distribution. 

    We used the $\chi^2$ statistics to estimate the best-fit parameters. We modelled the complex in the mHz range with two Lorentzians, to better describe both the broad shoulder at higher frequencies and the sharper peak at lower frequencies. To model the whole PDS we also considered a constant in order to account for the white noise component that dominates at frequencies $\nu\gtrsim0.1$\,Hz, and a power-law for the red noise component dominant at frequencies $\nu\lesssim1\E{-5}$\,Hz. The final model used to fit the PDSs is described by the following equation:
    \begin{equation}\label{eq:PDSmodel}
        P(\nu)=\mathrm{const}_\mathrm{WN}+K_\mathrm{RN}\nu^{\Gamma_\mathrm{RN}}+\sum_{i=1}^2K_i\frac{\Delta\nu_i}{2\pi}\frac{1}{(\nu-\nu_i)^2+(\Delta\nu_i/2)^2}
    \end{equation}
    where $P(\nu)$ is the power $P$ at the frequency $\nu$, the first two terms on the right-hand side describe the white and red noise, respectively, and the summation describes the two Lorentzians. $\nu_i$ is the centroid frequency of the Lorentzian and $\Delta\nu_i$ its full width at half maximum (FWHM). In all our fits we found that the white-noise constant and the red-noise, power-law index are consistent with the expected values (using the Leahy normalization) of $\mathrm{const}_\mathrm{WN}=2$ and $-2\lesssim\Gamma_\mathrm{RN}\lesssim-1$, respectively \citep[see e.g.][]{vanderKlis1989}.

    To estimate the significance of the Lorentzians we used two different, independent methods: the F-test as implemented in \textsc{XSPEC} (see e.g. \citealt{Strohmayer2007,Strohmayer2009}), and a procedure similar to the one described in \cite{Motta2015}. For the latter, we converted the PDSs to the square fractional rms normalization \citep{Belloni1990} and computed the integral of the power of each Lorentzian. In \cite{Motta2015}, they computed the ratio of the integral over the associated negative 1$\sigma$ error (see footnote 6 in the original work). To take into account any possible non-Gaussianity in the errors, for each Lorentzian component we computed the negative N$\sigma$ error for which each integral was consistent with 0. We will first describe the results obtained with the F-test and then those obtained with the N$\sigma$ method.

    For the F-test, we started from a model with only white and red noise. We found that this model results in unacceptable fits for every observation. We then added a Lorentzian to model the broad shoulder and computed its significance. We found that the fit strongly requires the broad shoulder, with a 4.8$\sigma$, 5.5$\sigma$, and 6.5$\sigma$ significance in observation A, B, and C, respectively. With the available data, we cannot exclude that the broad feature is actually an unresolved sum of harmonics, given the highly non-sinusoidal shape of the modulation \citep[see e.g.][]{Angelini1989}. The inclusion of a second Lorentzian to model the sharper peak with respect to the previous model is not strictly required, with a 2.9$\sigma$, 2.5$\sigma$ and 2.8$\sigma$ significance in observation A, B, and C, respectively. However, we note that if we add the second Lorentzian we consistently obtain better residuals at $\simeq0.5$\,mHz (see lower panels of Fig.~\ref{fig:psds}).

    It is known that the F-test may provide incorrect results in the case of lines \citep{Protassov2002}. By using the integral approach described above, we found that both Lorentzians (the broad shoulder and the sharper peak) are significant. The broad shoulder is detected with a 3.6$\sigma$, 5.0$\sigma$, and 4.3$\sigma$ significance in observation A, B, and C, respectively. The sharper peak is detected at similar confidence levels, with a 3.6$\sigma$, 3.2$\sigma$, and 4.3$\sigma$ significance in observation A, B, and C, respectively. Thus, we conclude that we detect both features with a significance $\ge3\sigma$.

    The results of our fits regarding the two Lorentzians, together with the corresponding rms fractional variability, are shown in Tab.~\ref{tab:PDSsfits_LP} for the three observations (A, B, and C) separately. Given the little variability shown by the Lorentzians among the three observations, we also fitted the three PDSs simultaneously (A+B+C).
    A feature in the PDS is usually defined as a QPO when its quality factor $Q=\nu/\Delta\nu$ is $Q\gtrsim2$. In our case, the low-frequency feature centred at $\nu_\mathrm{QPO}\simeq0.5$\,mHz can be classified as a QPO, since $Q_\mathrm{QPO}\gtrsim2$ in every observation (and up to 10). The high-frequency feature, on the other hand, always has $Q_\mathrm{broad}<2$, therefore we refer to this component as a broad shoulder, or simply as a broad feature. The rms fractional variability of each component is rms$_\mathrm{QPO}\simeq30\%$ and rms$_\mathrm{broad}\simeq40\%$, respectively, and appears stable among the observations. In Tab.~\ref{tab:PDSsfits_LP} we also reported for each component the characteristic frequency $\nu_\mathrm{char}=\sqrt{\nu^2+(\Delta\nu/2)^2}$ \citep[see e.g. Section 2 of][]{Ingram2019}. 
    Although the flare-like activity could lead us to describe the feature as quasi-periodic flarings (QPFs), we opted for the terminology used for other ULXs and classified them as QPOs.
    
    To study the energy evolution of these features we computed the PDSs in the soft (0.3--1.5\,keV) and hard (1.5--10\,keV) band. We report the results of our fits in Tab.~\ref{tab:PDSsfits_LP}. In the case of observation A, we adopted a geometric rebin of 30\%, due to poorer statistics (this observation is highly affected by particle flares). 
    The values found for the different parameters are consistent among the different energy bands. 

   \begin{table*}
       \caption{Parameters of the Lorentzians obtained from the fit of the 0.3--10\,keV, 0.3--1.5\,keV and 1.5---10\,keV PDSs of our LP observations. $\nu_x$: centroid frequency of the Lorentzian $x$. $\Delta\nu_x$: full width at half maximum (FWHM) of the Lorentzian $x$. $\nu_{\mathrm{char},x}$: characteristic frequency of the Lorentzian $x$. $Q_x=\nu_x/\Delta\nu_x$: approximate quality factor of the Lorentzian $x$. rms$_x$: rms fractional variability of the Lorentzian $x$. Errors given at 1$\sigma$ (68.3\%) confidence level. Two additional components have been added to the fit to take into account the white and red noise at high and low frequencies, respectively.
       }
       \centering
       \resizebox{\textwidth}{!}{
       \begin{tabular}{cccccccccccc}
       \hline\hline
          ObsID & $\nu_\mathrm{QPO}$ & $\Delta\nu_\mathrm{QPO}$ & $\nu_\mathrm{char,QPO}$ & $Q_\mathrm{QPO}$ & rms$_\mathrm{QPO}$ & $\nu_\mathrm{broad}$ & $\Delta\nu_\mathrm{broad}$ & $\nu_\mathrm{char,broad}$ & $Q_\mathrm{broad}$ & rms$_\mathrm{broad}$ & $\chi^2$/dof\\
          & (mHz) & (mHz) & (mHz) & & (\%) & (mHz) & (mHz) & (mHz) & & (\%) & \\
        \hline
        \multicolumn{12}{c}{0.3--10\,keV}\\
        \hline
        \\
         A & $0.449\apm{0.022}{0.019}$ & $0.088\apm{0.035}{0.054}$ & $0.451\apm{0.022}{0.019}$ & $5.1$ & $29.0\apm{4.0}{3.8}$ & $1.20\apm{0.27}{0.26}$ & $2.60\apm{0.53}{0.67}$ & $1.77\apm{0.27}{0.30}$ & $0.5$ & $37.9\apm{2.6}{2.7}$ & 27.74/35\\ \\
         B & $0.470\apm{0.017}{0.012}$ & $0.046\apm{0.046}{0.053}$ & $0.470\apm{0.017}{0.011}$ & $10.2$ & $27.4\pm4.3$ & $0.92\apm{0.14}{0.18}$ & $1.65\apm{0.26}{0.31}$ & $1.24\apm{0.13}{0.17}$ & $0.6$ & $38.6\apm{2.7}{2.6}$ & 24.23/35 \\ \\
         C\tablefootmark{a} & $0.519\apm{0.033}{0.036}$ & $0.183\apm{0.061}{0.069}$ & $0.527\apm{0.033}{0.036}$ & $2.8$ & $32.0\pm3.6$ & $1.56\apm{0.23}{0.25}$ & $2.74\apm{0.43}{0.53}$ & $2.08\apm{0.22}{0.26}$ & $0.6$ & $40.3\pm2.6$ & 46.66/37 \\ \\
         A+B+C & $0.565\apm{0.036}{0.034}$ & $0.269\apm{0.054}{0.067}$ & $0.581\apm{0.035}{0.034}$ & 2.1 & $29.5\pm2.4$ & $1.34\pm0.17$ & $2.45\apm{0.31}{0.37}$ & $1.81\apm{0.16}{0.18}$ & 0.5 & $36.1\pm1.8$ & 128.31/124\\ \\
         \hline
            \multicolumn{12}{c}{0.3--1.5\,keV}\\
        \hline
        \\
        A\tablefootmark{b} & $0.534\apm{0.027}{0.024}$ & $0.148\apm{0.081}{0.062}$ & $0.539\apm{0.028}{0.025}$ & 3.6 & $32.9\apm{4.0}{3.4}$ & $1.48\apm{0.20}{0.18}$ & $1.56\apm{0.53}{0.76}$ & $1.67\apm{0.22}{0.24}$ & 0.9 & $32.3\apm{3.4}{2.5}$ & 32.26/24 \\ \\
        B & $0.467\apm{0.017}{0.014}$ & $0.061\apm{0.035}{0.052}$ & $0.468\apm{0.017}{0.014}$ & 7.6 & $29.7\apm{4.5}{4.2}$ & $1.04\apm{0.18}{0.40}$ & $1.21\apm{0.48}{0.76}$ & $1.20\apm{0.19}{0.39}$ & 0.9 & $29.7\apm{3.5}{3.7}$ & 43.13/35 \\ \\
        C & $0.484\apm{0.028}{0.031}$ & $0.184\apm{0.063}{0.058}$ & $0.493\apm{0.028}{0.031}$ & 2.6 & $31.8\apm{3.6}{3.5}$ & $1.54\apm{0.24}{0.21}$ & $1.81\apm{0.42}{0.43}$ & $1.79\apm{0.24}{0.21}$ & 0.9 & $33.8\pm3.2$ & 25.33/35 \\ \\
         \hline
            \multicolumn{12}{c}{1.5--10\,keV
            }\\
        \hline
        \\
        A\tablefootmark{b} & $0.509\apm{0.044}{0.072}$ & $0.25\apm{0.10}{0.22}$ & $0.525\apm{0.045}{0.075}$ & 2.0 & $36.9\apm{5.5}{5.1}$ & $1.52\apm{0.43}{0.26}$ & $1.21\apm{0.63}{0.66}$ & $1.64\apm{0.42}{0.27}$ & 1.3 & $32.7\apm{6.3}{7.6}$ & 34.30/23 \\ \\
         B & $0.469\apm{0.022}{0.014}$ & $0.047\apm{0.047}{0.067}$ & $0.470\apm{0.022}{0.014}$ & 9.9 & $25.8\pm5.3$ & $1.03\apm{0.18}{0.19}$ & $1.68\apm{0.31}{0.38}$ & $1.33\apm{0.17}{0.19}$ & 0.6 & $46.3\apm{3.3}{3.8}$ & 22.67/35 \\ \\
         C & $0.538\apm{0.041}{0.028}$ & $0.26\apm{0.10}{0.12}$ & $0.553\apm{0.042}{0.030}$ & 2.1 & $34.6\pm5.4$ & $1.27\apm{0.53}{0.71}$ & $4.08\apm{0.81}{0.98}$ & $2.40\apm{0.44}{0.55}$ & 0.3 & $47.5\apm{4.9}{3.9}$ & 34.71/35 \\ \\
        \hline
       \end{tabular}
       }
        \tablefoot{
        \tablefoottext{a}{Due to poor statistics at low frequencies, we froze red-noise parameters at the best-fit values estimated before the addition of the two Lorentzians.}
        \tablefoottext{b}{Geometric rebin changed to a factor 30\%, due to poor statistics.}
        }
       \label{tab:PDSsfits_LP}
   \end{table*}

    We looked for the presence of the 2.8-s spin signal by using acceleration algorithms and also including a first-period derivative component (see \citealt{RodriguezCastillo2020} for more details), but we found no peak associated with a periodic signal, within a reasonable period interval (see below the range for $P_\mathrm{exp}$), in any of the three PDSs. To derive an upper limit on the pulsed fraction PF of the spin signal, we started from the timing solutions of \cite{Brightman2022} ($P=2.78674(4)$\,s, the last time \ulxie\ spin signal was detected, on 2019 July 12) and \cite{RodriguezCastillo2020} ($P\simeq2.79771475(25)$\,s, on 2018 June 14). From these values we derived a secular spin period derivative $\dot{P}\simeq-3.23(1)\E{-10}$\,s\,s$^{-1}$. In the search we considered a conservative value of $\left|\dot{P}\right|<10^{-8}$\,s\,s$^{-1}$. The expected spin signal period is, therefore, in the $2.78458<P_\mathrm{exp}<2.7889$\,s.  
    Finally, we followed the procedure described by \cite{Israel1996} to compute the frequency-dependent detection threshold and the 3$\sigma$ upper limit on PF$_\mathrm{range}$ in the 0.350--0.374\,Hz range (equivalent to the 3$\sigma$ confidence range on $\nu_\mathrm{exp}$), obtaining PF$_\mathrm{range}\lesssim10\%$. The single trial PF upper limit on a sinusoidal signal at $\nu_\mathrm{exp}$ is PF$_\mathrm{single}\lesssim6\%$. 

    We then searched for other occurrences of this mHz-complex in the PDSs of the archival X-ray observations we selected according to the criteria outlined in Sect.~\ref{sec:ObsDataReduction}.
    The timing analysis of \nustar\ observation 60501023002 was already performed by \cite{Brightman2022}: neither \ulxie\ spin signal nor the mHz-QPOs were detected during this observation. 

    \begin{figure}
        \centering
        \includegraphics[width=\columnwidth]{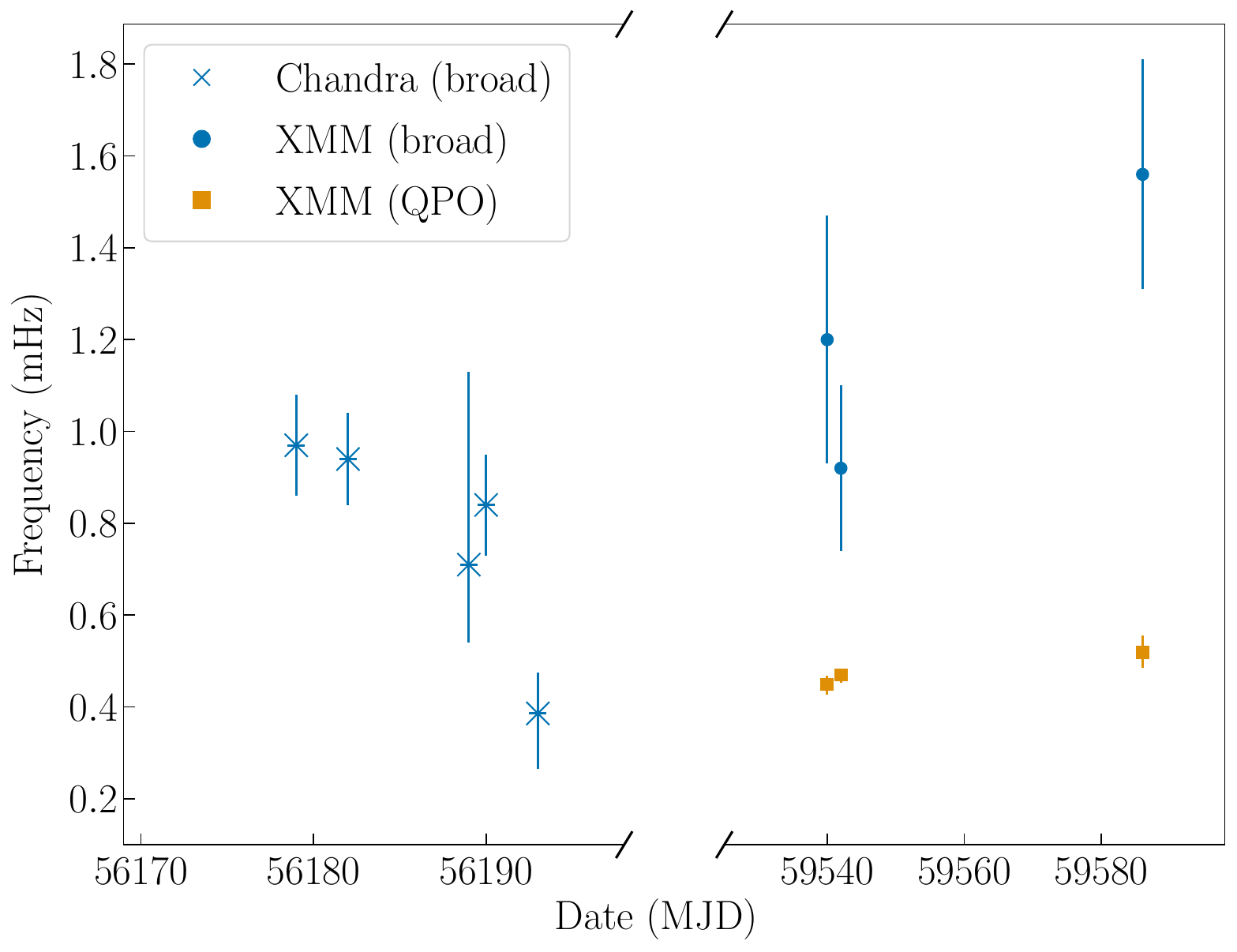}
        \includegraphics[width=\columnwidth]{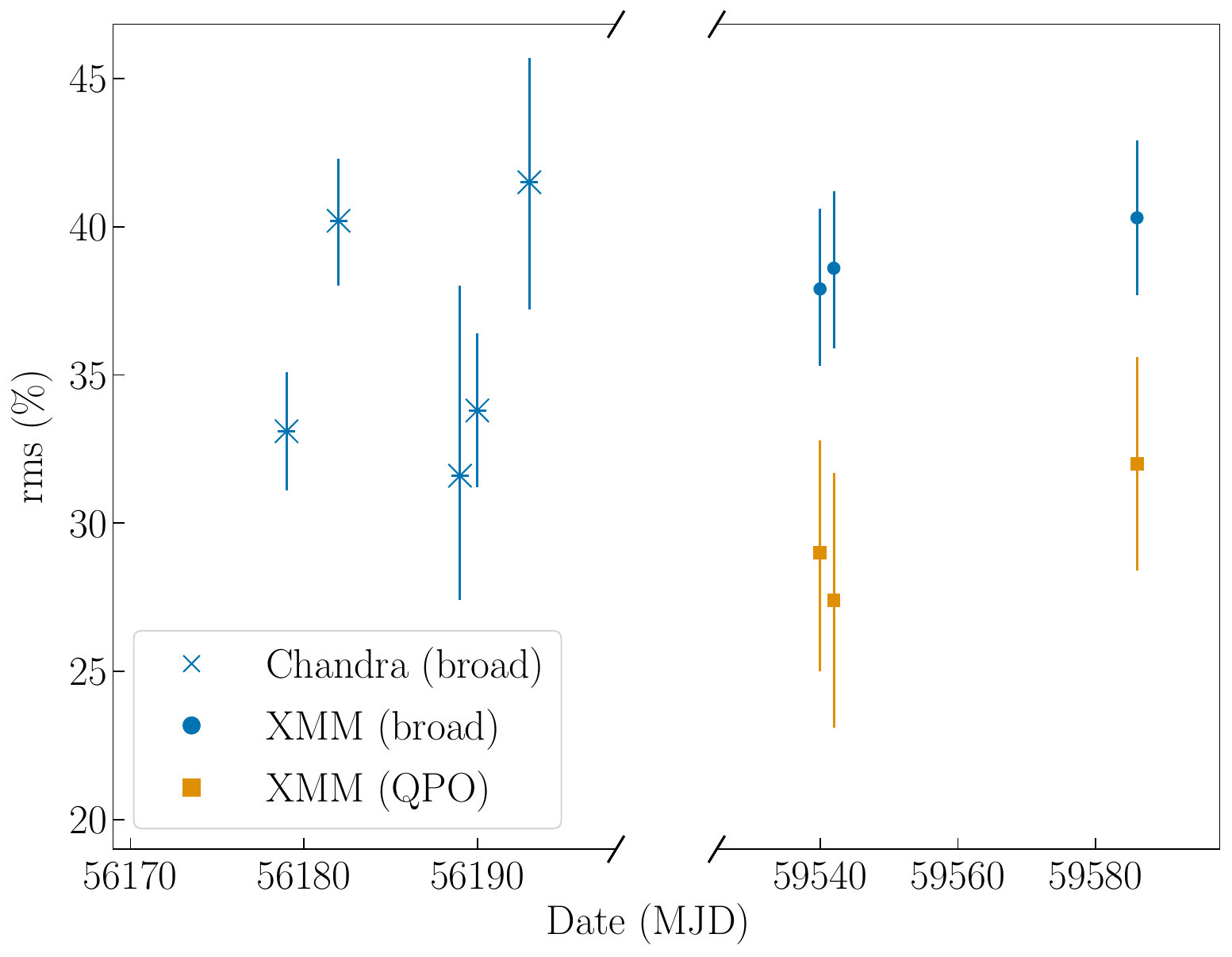}
        \caption{Top panel: evolution of the centroid frequency $\nu$. Bottom panel: evolution of the fractional rms. Cross markers represent the values for the broad feature in \chandra\ data, while blue circles and orange squares represent the values for the broad feature and the candidate QPO, respectively, in the latest \xmm\ observations. For \chandra\ we considered events in the 0.5--10\,keV band, while for \xmm\ events in the 0.3--10\,keV band.}
        \label{fig:QPOevo}
    \end{figure}

    In Tab.~\ref{tab:XrayObs} we highlighted the observations for which the complex at mHz range has a significance $\ge3\sigma$. We found 5 consecutive \chandra\ observations, performed between 2012 September 9 and 2012 September 23, during which it was detected. We report the details of our analysis of the \chandra\ observations in Appendix~\ref{appendix:Chandra}. For these observations, only one Lorentzian was required by the fit. Given the low quality factor we derive for the Lorentzian (see Tab.~\ref{tab:Chandra_PDSs}), we associated it with the broad feature. In Fig.~\ref{fig:QPOevo} we show the evolution of the centroid frequency and the fractional rms of the broad feature and the QPO. Both the frequency and the rms show little variability between the two epochs, albeit separated by almost 10 years.

    \begin{figure}
        \centering
        \includegraphics[width=\columnwidth]{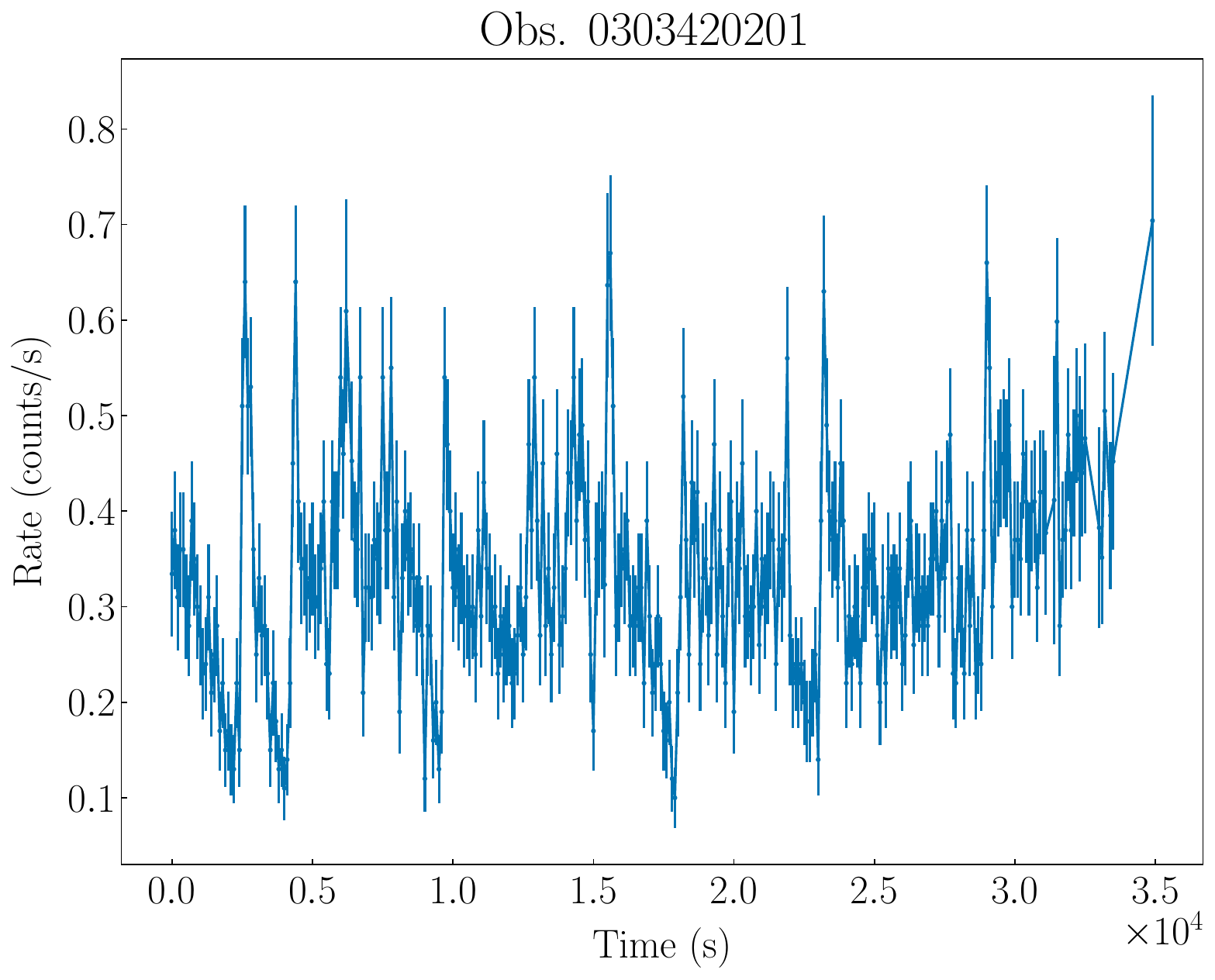}
        \includegraphics[width=\columnwidth]
        {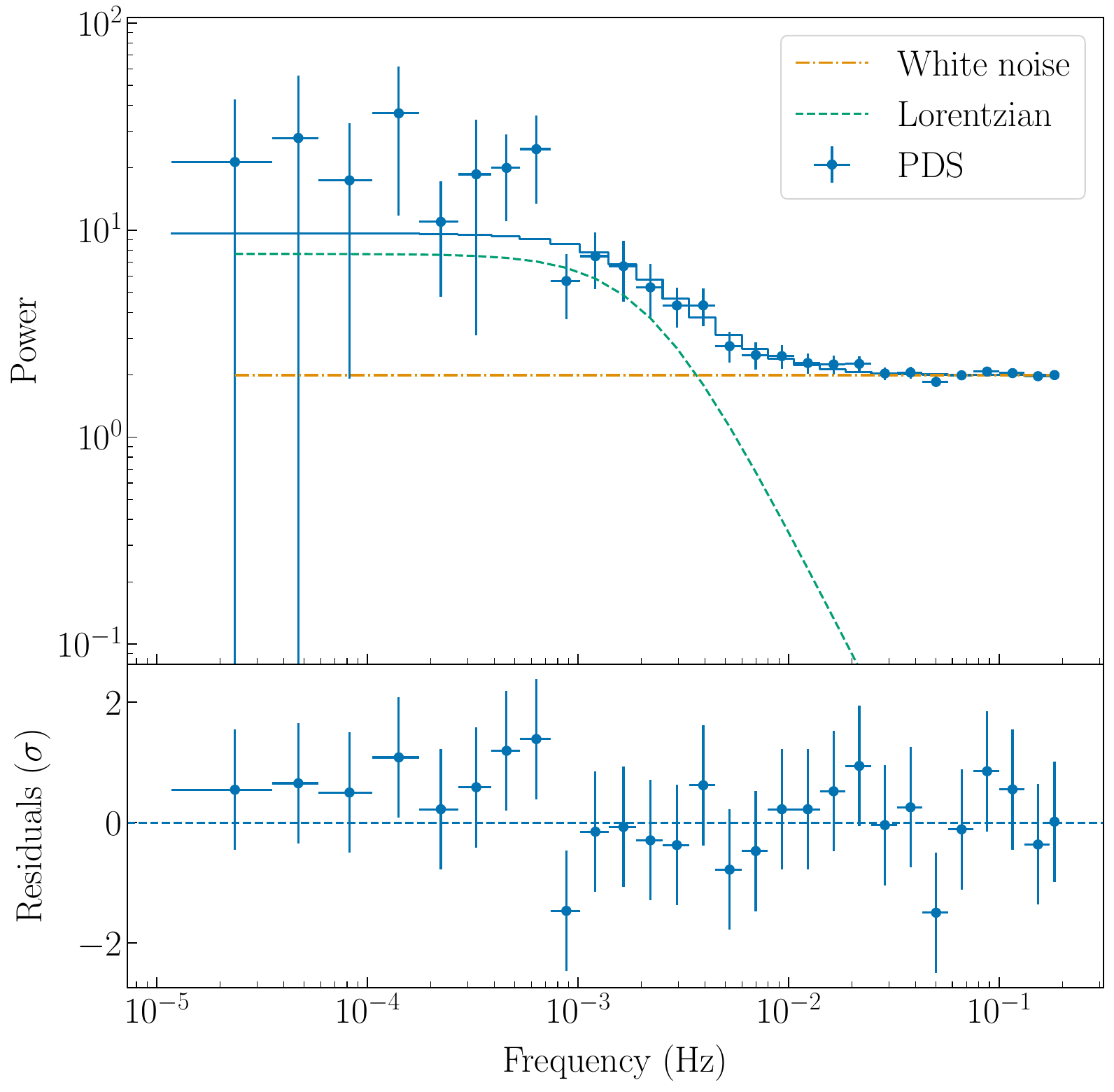}
        \caption{Top figure: PN+MOS light curve of \ulxie, \xmm\ observation \obsidold. The bin time is 100\,s, as in the three LP light curves. Bottom figure: PDS in the 0.3--10\,keV band of \xmm\ observation \obsidold\ (top panel) and residuals of the white noise plus single Lorentzian model used for this observation (bottom panel). We combined data from PN and MOS cameras. The PDS is computed using the Leahy normalization.}
        \label{fig:XMMOld}
    \end{figure}

    We also confirmed a different type of variability, first reported by \cite{Earnshaw2016}, in an archival \xmm\ observation (ObsID \obsidold), for which we show the 0.3--10\,keV light curve and the associated PDS with a logarithmic rebin factor of 1.32 in the top and bottom panel of Fig.~\ref{fig:XMMOld}, respectively. In this case, we found that red noise at the low-frequency end of the PDS was not required by the fit. Instead, we fitted the low-frequency tail of the PDS with a Lorentzian centred at 0, finding a width $\Delta\nu=4.2\apm{1.1}{1.2}$\,mHz ($\chi^2$/dof$=14/24$).

    \subsection{Spectral Analysis}\label{sec:LPSpectral}
    
    We analysed \ulxie\ spectra from observations B and C with the \textsc{XSPEC} package \citep{Arnaud1996} also used for the fitting of the PDSs. We stress again that we excluded observation A for our spectral analysis due to particularly high particle flare contamination. For the computation of the absorption column densities we adopted the element abundances and cross sections of \cite{Wilms2000} and \cite{Verner1996}, respectively. The uncertainties reported for the parameters of the considered spectral models represent a 90\% confidence range. We computed the absorbed and unabsorbed fluxes in the 0.3--10\,keV band using the pseudo-model \textsc{cflux}. To derive the unabsorbed luminosities we considered a distance from the source of 8.58\,Mpc \citep{McQuinn2016}.

    We modelled the spectra with two absorbed multi-temperature disk black bodies (following the approach of previous works, e.g. \citealt{Gurpide2021} and \citealt{Brightman2022}) using the \textsc{XSPEC} component \textsc{diskbb} \citep{Mitsuda1984}. For the absorption, we considered two separate \textsc{tbabs} components for the Galactic and intrinsic column densities. The former was fixed to $3.3\times10^{20}$ cm$^{-2}$ \citep[][]{HI4PICollaboration2016}, while the latter was left free to vary. At first, we fitted the two spectra separately and found the best-fit parameters consistent with each other, indicative of the same spectral state. Hence, we fitted them simultaneously, to increase the precision on the best-fit parameters. This latter fit resulted in a $\chi^2/\mathrm{d.o.f.}=607.21/651$ and a null hypothesis probabilty n.h.p.$=0.889$, with best-fit temperatures of the two disks of 0.3\,keV and $\sim$
    2.7\,keV, respectively. All best-fit results are reported in Table \ref{tab:BestFitXMM}. Both temperatures are consistent with those found for other ULXs \citep[e.g.,][]{Gurpide2021}. A check for intercalibration issues between the cameras, with a multiplicative renormalisation constant in the model, led to non-significant differences in the best-fit parameters, nor in the goodness of the fit.
    
\begin{figure}
        \centering
        \includegraphics[width=\columnwidth]{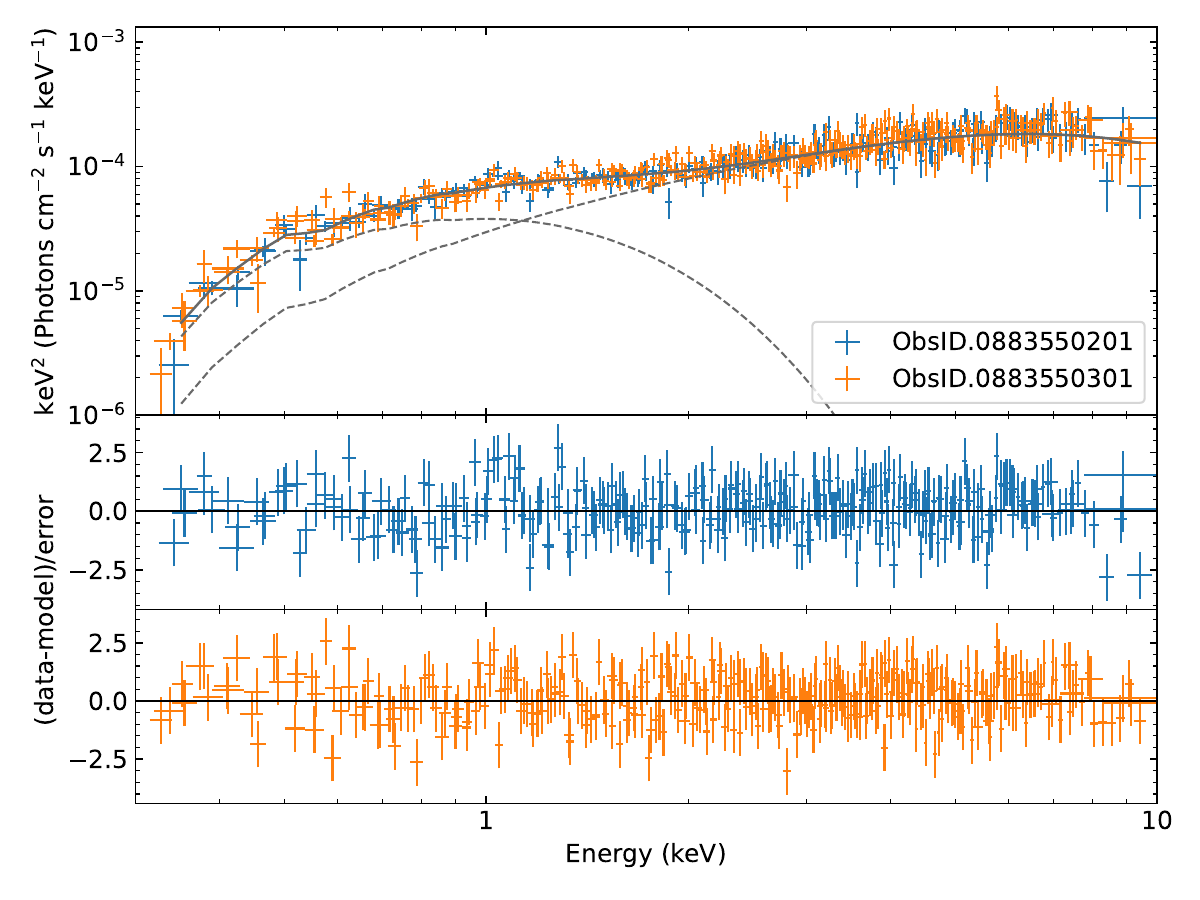}
        \caption{Top panel: Simultaneous fit of the two PN+MOS spectra from observations B (blue) and C (orange) with the double-disk model. Bottom panels: Residuals of the fit in units of standard deviation. Spectra of each colour include PN, MOS1 and MOS2 data (not summed) from the same observation. 
        }
        \label{fig:FitSpectraResiduals}
    \end{figure}

    We tested the hypothesis of a third spectral component at higher energies, as observed in some ULXs \citep[see e.g.][]{Walton2018}. Following \cite{Brightman2022}, who analysed simultaneous \xmm+\nustar\ observations of \ulxie, we added a cutoff power law to the spectral model (\textsc{cutoffpl}). Due to the lack of data at energies above 10 keV, we froze the power law photon index and cutoff energy to the best-fit values of \citet{Brightman2022} of $\Gamma=0.8$ and $E_\mathrm{cut}=8.1$\,keV, respectively. This new fit did not significantly improve the statistics compared to the previous one ($\chi^2/\mathrm{d.o.f.}=607.21/650$ and n.h.p.$=0.884$), with an upper limit on the power law normalisation of $8\times10^{-6}$ \,photons\,keV$^{-1}$\,cm$^{-1}$\,s$^{-1}$.
    Hence, we decided to keep just the two thermal components. \cite{RodriguezCastillo2020} reached an identical conclusion for previous observations of \ulxie\ when only \xmm\ data were available.  
    In general, the cutoff power law is needed only when data above 10\,keV are available, and in the case of a NS accretor this component is associated with the emission from the accretion column \citep[see e.g.][]{Walton2018}. We report the spectra of observations B and C, together with the double-disk model resulting from the simultaneous fit of the two observations, in Fig.~\ref{fig:FitSpectraResiduals}. 

    For purpose of comparison with \cite{RodriguezCastillo2020}, we also fitted the spectra with a black body spectrum (\textsc{bbodyrad}) in place of the harder multi-temperature black body component. Also in this case, we first verified that the best-fit parameters were consistent between the two observations and successively we fitted tying all of them together. We obtained a intrinsic absorption of $(5.2\apm{1.2}{1.3})\times10^{20}$\,cm$^{-2}$, a disk black body temperature of $0.48\pm0.03$\,keV and a black body temperature of $1.47\pm0.05$\,keV. The best-fit statics are $\chi^2/\mathrm{d.o.f.}=609.66/651$ and n.h.p.$=0.875$. All values are consistent with the results of the phase-resolved spectroscopy of  \cite{RodriguezCastillo2020}, but specifically, the black body temperatures are more consistent with the phases of minimum and raise/decay of the source. 

    We also noticed marginal evidence for excess residuals at about 1\,keV during observation B (see Fig.\ref{fig:FitSpectraResiduals}, middle panel). Those residuals are common to ULXs and are interpreted as blended, unresolved spectral lines caused by disk outflows at fractions of the speed of light \citep[see e.g.][]{Middleton2015} thanks to their unambiguous identification in high-resolution X-ray spectra \citep[see e.g.][]{Pinto2016,Pinto2023}.  

    To check for significant spectral differences between different phases of the aperiodic modulation, we extracted the spectra in two different intensity intervals. We defined a ``peak'' phase whenever the background-subtracted count rate was higher than 0.2 and 0.07\,counts\,s$^{-1}$ in the PN and MOS1/2 camera, respectively, and a ``no-peak'' phase whenever the count rate was lower. We chose these values since they are a good match to the plateau which can be seen sometimes between a peak and the subsequent minimum in the light curve. 
    
    We fitted the spectra of the observations with the double-disk model used above, first individually and then together. 
    The results of our fits are reported in Tab.~\ref{tab:BestFitXMMphase}. We found no significant difference in the spectral shape, with the best-fit parameters being consistent with each other within the error bars between the two phases. There is a systematic shift in the normalisations, especially in the soft disk component, albeit barely significant, but this is expected given the choice of the phases based on the count rate. We also computed the unabsorbed flux in the 0.3--10\,keV band of the soft and hard disk components separately and in both the peak and no-peak phases. We derived $F_\mathrm{peak}^\mathrm{soft}=(1.77\pm0.07)\E{-13}\ergcms$ and $F_\mathrm{no-peak}^\mathrm{soft}=(1.01\pm0.03)\E{-13}\ergcms$ for the soft disk in the two phases, while for the hard disk we derived $F_\mathrm{peak}^\mathrm{hard}=(6.68\pm0.16)\E{-13}\ergcms$ and $F_\mathrm{no-peak}^\mathrm{hard}=(4.12\pm0.07)\E{-13}\ergcms$. Both components increase their flux in the peak phase of a factor $\sim$1.7, suggesting that the overall shape of the spectrum remains unaltered. 

    \begin{table*}
      \caption[]{Best-fit spectral parameters of the latest \xmm{} observations with the double-disk model. 
      }
         \label{tab:BestFitXMM}
     \resizebox{\textwidth}{!}{
         \begin{tabular}{lcccccccccc}
         \hline\hline\noalign{\smallskip}
         Observation & $n_\mathrm{H}$\tablefootmark{a} & $kT_\mathrm{soft}$ & Norm. & $kT_\mathrm{hard}$ & Norm. & Flux\tablefootmark{b} & Lum.\tablefootmark{c} & $\chi^2$/dof & n.h.p.  \\
         \noalign{\smallskip}
         & ($10^{20}\,\mathrm{cm}^{-2}$) & (keV) & & (keV) & ($10^{-4}$) & ($10^{-13} \ergcms$) & ($10^{39} \ergs$) & \\
            \noalign{\smallskip}
            \hline
            \noalign{\smallskip}
            B & $9.1\apm{2.7}{3.1}$ & $0.32\apm{0.03}{0.04}$ & $0.7\apm{0.3}{0.6}$ & $2.63\apm{0.17}{0.20}$ & $5.7\apm{1.3}{1.5}$ & $5.37\pm0.08$ & $5.34\pm0.08$ & 297.93/309 & 0.664\\
            \noalign{\smallskip}
            C & $8.1\apm{2.3}{2.5}$ & $0.33\pm0.03$ & $0.6\apm{0.2}{0.4}$ & $2.78\apm{0.17}{0.21}$ & $4.6\apm{1.0}{1.2}$ & $5.37\pm0.07$ & $5.31\pm0.07$ & 306.91/337 & 0.879 \\
            \noalign{\smallskip} 
            \hline
            \noalign{\smallskip}
            B$+$C & $8.5\apm{1.8}{1.7}$ & $0.33\pm0.02$ & $0.6\apm{0.2}{0.3}$ & $2.71\apm{0.12}{0.13}$ & $5.0\apm{0.8}{0.9}$ & $5.37\pm0.05$ & $5.33\pm0.05$ & 607.21/651 & 0.889 \\
            \noalign{\smallskip}
            \hline\hline
         \end{tabular}}
     \tablefoot{
     \tablefoottext{a}{The Galactic absorption component was fixed to $n_\mathrm{H,gal} = 3.3\E{20}\nH$ \citep{HI4PICollaboration2016}.}
     \tablefoottext{b}{Observed flux in the 0.3--10\,keV band.}
     \tablefoottext{c}{Unabsorbed luminosity in the 0.3--10\,keV band.}
     }
   \end{table*}

     \begin{table*}
      \caption[]{Best-fit parameters of the spectra during the peaks and the minima (no-peak) of the modulation of the latest \xmm{} observations. We considered the same double-disk model as before. 
      }
         \label{tab:BestFitXMMphase}
      \resizebox{\textwidth}{!}{
         \begin{tabular}{lcccccccccc}
         \hline\hline\noalign{\smallskip}
         Observation & $n_\mathrm{H}$\tablefootmark{a} & $kT_\mathrm{soft}$ & Norm. & $kT_\mathrm{hard}$ & Norm. & Flux\tablefootmark{b} & Lum.\tablefootmark{c} & $\chi^2$/dof & n.h.p.  \\
         \noalign{\smallskip}
         & ($10^{20}\,\mathrm{cm}^{-2}$) & (keV) & & (keV) & ($10^{-4}$) & ($10^{-13} \ergcms$) & ($10^{39} \ergs$) & \\
            \noalign{\smallskip}
            \hline
            \noalign{\smallskip}
            B \\
            peak & $10.0\apm{4.7}{5.8}$ & $0.31\apm{0.05}{0.07}$ & $1.1\apm{0.7}{2.0}$ & $2.7\apm{0.3}{0.4}$ & $7.0\apm{2.7}{3.6}$ & $7.4\pm0.2$ & $7.4\pm0.2$ & 159.28/162 & 0.546\\
            \noalign{\smallskip}
            no-peak & $7.8\apm{3.1}{3.5}$ & $0.34\apm{0.04}{0.05}$ & $0.5\apm{0.3}{0.5}$ & $2.8\pm0.3$ & $4.0\apm{1.3}{1.7}$ & $4.67\pm0.09$ & $4.61\pm0.09$ & 288.73/255 & 0.072 \\
            \noalign{\smallskip}
            C \\
            peak & $10.8\apm{4.3}{5.1}$ & $0.30\apm{0.04}{0.05}$ & $1.4\apm{0.8}{1.9}$ & $2.9\apm{0.3}{0.4}$ & $5.7\apm{2.1}{2.8}$ & $7.30\pm0.18$ & $7.46\pm0.18$ & 175.12/180 & 0.589 \\
            \noalign{\smallskip}
            no-peak & $5.8\apm{2.6}{3.0}$ & $0.36\apm{0.04}{0.05}$ & $0.32\apm{0.15}{0.28}$ & $2.8\apm{0.2}{0.3}$ & $4.0\apm{1.1}{1.4}$ & $4.62\pm0.08$ & $4.45\pm0.08$ & 277.37/277 & 0.482 \\
            \noalign{\smallskip} 
            \hline
            \noalign{\smallskip}
            B$+$C \\
            peak & $10.5\apm{3.2}{3.7}$ & $0.31\apm{0.03}{0.04}$ & $1.3\apm{0.6}{1.2}$ & $2.8\apm{0.2}{0.3}$ & $6.2\apm{1.7}{2.1}$ & $7.35\pm0.13$ & $7.45\pm0.13$ & 337.69/347 & 0.63 \\
            \noalign{\smallskip}
            no-peak & $6.6\apm{2.0}{2.2}$ & $0.35\apm{0.03}{0.04}$ & $0.39\apm{0.14}{0.23}$ & $2.76\apm{0.17}{0.20}$ & $4.0\apm{0.8}{1.0}$ & $4.64\pm0.06$ & $4.52\pm0.06$ & 567.63/537 & 0.174\\
            \noalign{\smallskip}
            \hline\hline
         \end{tabular}
        } 
     \tablefoot{
     \tablefoottext{a}{The Galactic absorption component was fixed to $n_\mathrm{H,gal} = 3.3\E{20}\nH$ \citep{HI4PICollaboration2016}.}
     \tablefoottext{b}{Observed flux in the 0.3--10\,keV band.}
     \tablefoottext{c}{Unabsorbed luminosity in the 0.3--10\,keV band.}
     }
   \end{table*}

    The spectral analysis of the archival observations showing the broad variability feature has been already performed by \cite{Earnshaw2016}. Their results show that \ulxie\ has always been detected at a super-Eddington luminosity (few 10$^{39}\ergs$) in all those observations during which the broad feature in the PDS is detected. They also found that the spectrum of observation \obsidold\ (the one showing the different type of variability) can be well described ($\chi^2/\mathrm{dof}=172.3/182$) by a single power-law component with spectral index $\Gamma\simeq1.45$. The same model, when applied to observations A, B, and C, results in a worse reduced $\chi^2$, and structured residuals are visible, leading us to rule out the model. With the available data, however, we cannot assess a significant variation in the spectral state. We found that the spectrum in observation \obsidold\ can be described equally well ($\chi^2/\mathrm{dof}=166.9/179$) by our double disk model, with parameters consistent with the ones found in observations A, B, and C. Moreover, other models (such as a blackbody plus power-law model) work equally well for observation \obsidold, suggesting that the statistics are not robust enough to favour one model over the others.


\section{Discussion}\label{sec:Discussion}

The detection of a broad complex in the mHz-range of the PDSs of the 2021-2022 \xmm\ observations marks \ulxie\ as the second PULX showing QPOs at super-Eddington luminosities. Moreover, it is the first time this feature is detected in multiple observations over more than one month of baseline, suggesting that, when present, these QPOs are rather stable. By inspecting archival \chandra\ observations, spanning two weeks in 2012 (marked with a dagger in Tab.~\ref{tab:XrayObs}), we significantly (>$3\sigma$) detected QPOs with similar properties. It is interesting to note that, even though the two epochs are separated by more than 10 years, the QPOs seem to show very little variation in both the centroid frequency and the fractional rms. Furthermore, observation \obsidold\ is the only observation out of 25 during which a flat-top noise best describes the low-frequency variability centred at 0\,Hz \citep[see also][]{Earnshaw2016}. The flickering pattern shown by the light curve during this observation was recently noticed also by \cite{Kovacevic2022}, who suggested a quasi-periodic origin. However, with the available data, we cannot tell whether \ulxie\ was observed in a different state or the difference in the PDS arises from the shorter length of the observation (36\,ks, compared to our three 130\,ks-long observations). 

\cite{Feng2010} reported a similar (lack of) evolution in the QPO of M82 X-2. They detected a 3-mHz QPO in three different \chandra\ observations, performed in 2005, 2007, and 2008, respectively.  For our discussion, we do not consider the two \xmm\ observations analyzed by \cite{Feng2010} since, as already pointed out by the authors in their original work, \xmm\ does not have the angular resolution needed to resolve M82 X-2 and M82 X-1 and contamination is always present. For the same reasons, caution is needed when interpreting the tentative detection of 8-mHz QPOs in M82 X-2 with \xmm\ reported by \cite{Caballero-Garcia2013}. The centroid frequency $\nu_\mathrm{QPO}$, the FWHM $\Delta\nu$ and the fractional rms are all consistent with each other among the three \chandra\ observations analyzed by \cite{Feng2010}. The presence of a mHz-QPO with little to no evolution among different epochs, therefore, might represent an interesting property of the PULXs when at super-Eddington luminosities. Given the small sample, we emphasize that at the moment this is only a (tantalizing) hypothesis, to be confirmed by searching for QPOs in other PULXs. To draw a comparison with other super-Eddington pulsars, the Galactic PULX Swift\,J0243.6+6124 shows QPOs at $\simeq$30--40\,mHz only when in the sub-Eddington regime \citep{Wilson-Hodge2018,Chhotaray2024}. This fact could indicate that PULXs consistently observed at super-Eddington luminosities (such as M82 X-2 and \ulxie) and PULXs showing shorter outbursts at these luminosity levels (such as Swift\,J0243.6+6124) behave differently.

Although we restricted our analysis to \ulxie\ observations where QPOs have a significance >$3\sigma$, it is worth mentioning that other \chandra\ and \xmm\ observations show similar variability in both the light curves and the PDSs. The light curve of \chandra\ observation 354 shows two cycles of a modulation at $\simeq$7620\,s \citep{Liu2002}, later confirmed by \cite{Yoshida2010}, while \xmm\ observation 0112840201 shows a modulation at a different period ($\simeq$5900\,s, \citealt{Dewangan2005}). Besides, \chandra\ observation 23474 shows 4 cycles of a 10-ks long modulation. In all these observations the significance of the modulation is below the 3$\sigma$ threshold. For most of them, however, the low significance probably arises from the short length of the observations themselves (typically $\lesssim50$\,ks), during which very few cycles of the modulation can be probed. 6 out of 8 observations where QPOs were significantly detected are longer than 100\,ks. 

\subsection{Comparison with other sources and possible explanations}

The flare-like activity shown in the three light curves of Fig.~\ref{fig:lc_all_obs} is reminiscent of the heartbeat shown by 4XMM\,J111816.0–324910 \citep[J1118, see][]{Motta2020} and of the quasi-periodic ``whispers'' shown by 4XMM\,J140314.2+541806 \citep[J1403, see][]{Urquhart2022}. Both sources are ULXs and they both show QPOs at super-Eddington luminosities, the former at a centroid frequency $\nu_\mathrm{QPO}\simeq0.4$\,mHz and the latter at $\nu_\mathrm{QPO}\simeq1.5$\,mHz, but neither of them are known to host a pulsar. In the case of J1118, the PDS showed a structured feature, composed of different peaks at 0.3--0.7\,mHz, with a low-frequency shoulder at 0.2\,mHz. The striking similarity with the heartbeat shown by the Galactic BH binary GRS 1915+105 \citep[see the $\rho$ class variability in][]{Belloni2000} suggests that limit-cycle instability driven by Lightman-Eardley radiation pressure instability \citep{Lightman1974} is the source of the modulation. For J1403, several models have been proposed to explain the QPOs: Lense-Thirring precession of an outflow from the inner regions of the disk \cite[][M19 hereafter]{Middleton2019}; marginally stable He burning from matter accreted on different regions of the surface of the NS \citep{Heger2007}; Lightman-Eardley instability again; beating between the NS spin frequency and the Keplerian frequency ($\nu_\mathrm{QPO}=\nu_\mathrm{orb}-\nu_\mathrm{spin}$) at the inner radius of a disk extending down to the magnetospheric radius $R_\mathrm{m}$, smaller than the corotation radius $R_\mathrm{co}$ in order to accrete matter onto the NS. This last model was also proposed to explain the mHz-QPOs seen during super-Eddington flares of LMC X-4 \citep{Moon2001}. LMC X-4 is a high-mass X-ray binary (HMXB) that shows parameters similar to \ulxie, with a binary period $\approx$1.4\,d, hosting a NS spinning with a period of $\approx$13.5\,s, spin--up/--down phases at a rate up to $|\dot{P}|\leq 10^{-10}$ s~s$^{-1}$, and a superorbital period $\approx$30.5\,d \citep[see][and references therein]{Molkov2017,Moon2001,Urquhart2022}.

The Lightman-Eardley instability is hard to reconcile with the differences we see in \ulxie\ with respect to J1118. $\rho$ class variability is characterized by a more stable variability pattern than the one we see in the \ulxie\ light curves. Besides, we do not see the expected spectral variation in the different phases of the modulation. Our spectral analysis of \ulxie\ observations shows no clear evolution in the spectral parameters between the peaks and the valleys of the modulation, apart from a (barely significant) change in the absorption column $n_\mathrm{H}$, which is higher at the peak of the modulation. J1118, on the other hand, shows a clear evolution among the different phases of the modulation. A higher column density could be the sign of outflows/winds from a slim disk. The presence of winds in ULXs is well established \citep[see e.g.][]{Pinto2016,Pinto2023wind} and excess residuals at $\simeq1$\,keV during observation B can be interpreted as an unresolved wind \citep{Middleton2015b}. Similar winds from slim disks are also observed in Galactic sources at (super-)Eddington luminosities, such as V404 Cyg \citep{King2015,Motta2017}. In this scenario, the flares would correspond to quasi-periodic phases where the disk is puffed up and launching the wind, while the out-of-flare emission corresponds to phases when the disk is thinner and not outflowing. This sort of quasi-periodic behaviour would explain the flaring we see in the light curve and the presence of a broad shoulder accompanying the QPOs. It could also explain why, compared with J1118, the flares are less regular and show a higher degree of diversity among different cycles. A similar scenario was invoked to explain the quasi-periodic dipping observed on 5-10 ks timescales, i.e. matching our findings, in the soft/supersoft source NGC 247 ULX-1 \citep{Alston2021}.

Theoretical works on He burning models predicts higher QPOs frequencies ($\sim10$\,mHz) and a higher rms at lower energies, which we do not detect. Therefore, we will not discuss this possibility any further. Two possible alternative scenarios for the observed QPOs remain to be discussed: Lense-Thirring precession (M19) and the beat frequency model (BFM; see e.g. \citealt{Lamb1985,Angelini1989}). We first need to compute magnetospheric and corotation radius for \ulxie:
\begin{equation}\label{eq:rm}
    R_\mathrm{m}=3.3\E{7}\xi_{0.5} B_{12}^{4/7}L_{39}^{-2/7}R_6^{10/7}M_{1.4}^{1/7}\,\mathrm{cm}
\end{equation}
\begin{equation}\label{eq:rco}
R_\mathrm{co}=\left(\frac{GM}{\Omega^2}\right)^{1/3}\simeq1.5\E{8}\left(\frac{M}{\solarM}\right)^{1/3}P^{2/3}\,\mathrm{cm}
\end{equation}
where $P$ is the spin period of \ulxie\ in seconds, $B_{12}$ its magnetic dipolar field strength in units of 10$^{12}$\,G, $L_{39}$ its luminosity in units of 10$^{39}\ergs$, $R_6$ its radius in units of 10$^6$\,cm, and $M_{1.4}$ its mass in units of 1.4$\solarM$. $\xi$ is a parameter that takes into account the geometry of the accretion flow and in the case of an accretion disk is $\approx$0.5 \citep{Ghosh1979a,Wang1987,Campana2018}. From our spectral analysis, we know that the 0.3--10\,keV luminosity of \ulxie\ is $\LX\simeq5.3\E{39}\ergs$. We assume for the radius and mass $R_6\simeq M_{1.4}\simeq1$, which are considered typical values for a NS, and for the spin period $P\approx2.78$\,s. If we plug these values into eq.~\eqref{eq:rco}, we obtain 
$R_\mathrm{co}\simeq3.3\E{8}$\,cm. In the BFM, a QPO frequency of $\nu_\mathrm{QPO}\approx1$\,mHz would correspond to a disk (truncated by the magnetosphere) whose inner radius $R_\mathrm{in}$ is only slightly smaller than $R_\mathrm{co}$. For simplicity, we can assume $R_\mathrm{m}\approx R_\mathrm{in}\approx R_\mathrm{co}$. From eq.~\eqref{eq:rm}, we derive a magnetic dipolar field $B\simeq1.3\E{14}$\,G, inconsistent with previous estimates ($10^{12}\,\mathrm{G}\lesssim B\lesssim10^{13}\,\mathrm{G}$, see \citealt{RodriguezCastillo2020}). Another problem with the BFM scenario is that with $R_\mathrm{m}\approx R_\mathrm{co}$ we would expect frequent drops in the X-ray luminosity (which we do not detect in \ulxie), since even a small fluctuation in the accretion rate would lead to $R_\mathrm{m}>R_\mathrm{co}$ and therefore to the propeller regime. Additionally, a disk's inner radius so close to the corotation radius would require a high level of fine-tuning. 

For the Lense-Thirring scenario, on the other hand, we follow M19, particularly the values reported in their Table 1. In their work, the mHz-QPOs originated from a precessing inflow with period $P_\mathrm{inflow}$, and they can be linked with the period of a precessing wind $P_\mathrm{wind}$. The latter is set equal to the superorbital period we observe in various PULXs, including \ulxie, which shows a superorbital period of $\approx$44\,d \citep{Brightman2022}. We can scale the values reported in the first two columns of Table 1 of M19 by the observed superorbital period. By doing so, we find that the frequencies we observe for \ulxie\ are consistent with the ones expected for a NS-ULX with a high fraction of energy dissipated to launch the winds ($\epsilon=0.95$ case). \ulxie's QPO frequency could therefore arise from Lense-Thirring precession of the inner accretion flow. The detection of winds from \ulxie\ with a precessing period equal to the superorbital period could further strengthen the hypothesis, particularly if they show up mainly during the observations with QPOs rather than pulsations detected.  

Problems with this interpretation arise once we consider the temperature of the cold component of the disk ($T_\mathrm{sph}$ in M19) and the $P_\mathrm{wind}/P_\mathrm{inflow}$ ratio. According to M19, the two quantities can help constrain the magnetic field of the accreting NS powering the PULX (see Fig.~5 in their work). In brief, given $T_\mathrm{sph}$, the higher $P_\mathrm{wind}/P_\mathrm{inflow}$, the lower the magnetic field (see orange curve in Fig.~5 of M19). With $T_\mathrm{sph}=kT_\mathrm{soft}\simeq0.3$\,keV (from the spectral analysis of observations B and C) and $P_\mathrm{wind}/P_\mathrm{inflow}\simeq3800$ (assuming $P_\mathrm{wind}\simeq44$\,d and $P_\mathrm{inflow}\simeq1$\,ks), \ulxie\ would have an unrealistically low magnetic field $B\ll10^9$\,G. Whilst it is difficult to reconcile such a low field strength with the fact that we see pulsations from the NS, $P_\mathrm{wind}/P_\mathrm{inflow}$ is highly sensitive to assumed parameters related to the wind-launching \citep{Middleton2018, Middleton2019} which may yet permit higher field strengths as a solution.

    \subsection{QPOs and pulsed fraction of the spin signal}

One interesting feature of the 2021-2022 \xmm\ observations is the non-detection of \ulxie\ 2.8-s spin pulsations. For this part of the discussion, we will only consider \xmm\ observations, since \chandra\ time resolution (3.14\,s in the analysed observations) is not good enough to detect the \ulxie\ spin signal. PULXs are known to show transient pulsations, even within the same observation (see e.g. \citealt{Bachetti2020}, and observation B in Fig.~4 of \citealt{RodriguezCastillo2020}). Nevertheless, our analysis did not detect the spin pulsations in any intervals of the three observations. Given the high spin derivative $\dot{P}$ typical of PULXs, which can change from observation to observation, we could only compute the 3$\sigma$ upper limit on the PF considering the single observations. The single-trial value we derived of $\simeq$6\% is consistent with the minimum PF detected in 2018 by \cite{RodriguezCastillo2020}, suggesting that, if present, we should have detected the spin pulsations. One could argue that the lack of pulsations may be due to the source being in a different spectral state.
The 2021-2022 \xmm\ observations were all performed at the peak of the super-orbital modulation, and our spectral analysis in Sect.~\ref{sec:LPSpectral} confirms that we observed \ulxie\ in a similar state to the 2018 \xmm\ campaign. All the spectral parameters are consistent between the two epochs, with only a 6\% difference in the 0.3--10\,keV observed luminosity (with respect to 2018 pointing A having a similar flux level). Interestingly, the only difference between the two sets of observations is the presence of the QPOs in 2021-2022. In 2018 the PDSs showed no significant features at the mHz-range. Moreover, after comparison with Tab.~1 of \cite{RodriguezCastillo2020}, we conclude that the QPOs are not present in any of the \xmm\ observations in which the spin pulsations were detected; vice versa, when the QPOs were detected, we do not detect the spin signal. We therefore suggest that, whatever mechanism is responsible for the QPOs, is also responsible for a significant decrease in the pulsed fraction of the spin pulsations, but at the same time does not produce a significant change in the spectral state of the source. 

The detection of mHz-QPOs in both M82 X-2 and \ulxie\ has profound implications for the ULX population as a whole. First of all, it demonstrates that this feature is not exclusive to BH-powered ULXs. The derivation of the mass of the accreting BH from the frequency of the QPOs, therefore, must be treated with extreme caution. \cite{Heil2009} already pointed out that, if the ULX state is different from the sub-Eddington accretion state, mass estimates from the QPO frequencies and PDS features are unreliable. Similarly, \cite{Poutanen2007} had noted that QPOs at mHz-range are also known for Galactic BHs like Cygnus\,X-1 \citep{Vikhlinin1994} and GRS\,1915+105 \citep{Morgan1997}, weakening the association of mHz-QPOs in ULXs with type-C QPOs. Lastly, \cite{Middleton2011} reanalysed \xmm\ observations of NGC 5408 X-1 (one of the ULXs showing mHz-QPOs) and demonstrated that both timing and spectral analysis do not support the IMBH scenario proposed by \cite{Strohmayer2007}. Nevertheless, even after these works and the discovery of PULXs, the vast majority of works on ULXs showing QPOs assumes (IM)BH accretors \citep[see e.g.][and references therein]{Atapin2019,Majumder2023}.

If QPOs are indeed a PULX signature, they might represent an additional element to single out a candidate PULXs. It is interesting to note, for example, that among the ULXs showing QPOs, there is also IC 342 X-1 \citep{Agrawal2015}, later identified as a PULX candidate by \cite{Pintore2017} based on its hard energy spectrum. Another ULX that shows a similar combination is M74 X-1, with flaring activity in the light curve, high variability among different observations and mHz-QPOs in the PDS \citep{Krauss2005}. The downside is that, apparently, the presence of the QPOs is concurrent with a significant decrease in the pulsed fraction of the spin pulsations. The detection of the spin signal from a PULX is notoriously a difficult task, often involving the use of accelerated search techniques to compensate for the PDS loss of power (at the spin frequency) caused by the huge spin-up of these sources together with orbital Doppler effects \citep[see e.g. the analysis and discussion in][]{RodriguezCastillo2020}. A QPO could further hinder the process of detecting spin signals. 

\section{Conclusions}\label{sec:Conclusioni}

We have reported on the discovery of persistent QPOs in the mHz-range in three \xmm\ observations of \ulxielong\ performed in 2021--2022. Concurrently, we did not detect the 2.8\,s-long spin signal, with a 3$\sigma$ upper limit on the pulsed fraction $\mathrm{PF}\lesssim10\%$. These findings represent the second time QPOs are detected in a PULX at super-Eddington luminosities, the first being M82 X-2. We searched for other occurrences of the mHz-feature in \ulxielong\ archival observations of \xmm, \chandra\ and \nustar\ and found other 5 \chandra\ observations during which the mHz-feature is significantly ($>3\sigma$) detected. Our spectral analysis of the 2021--2022 \xmm\ observations shows that the source was observed in a similar state with respect to the 2018 \xmm\ observations when the spin pulsations were first detected \citep{RodriguezCastillo2020}.

We considered different models proposed to explain similar variability patterns in other ULXs. A disk puffing up and launching winds with a quasi-periodic recurrence could explain the flaring-like light curve and the broad feature in the PDS. Another viable explanations for the mHz-QPOs is Lense-Thirring precession of an outflow from the inner regions of the disk. We note, however, that the latter need a high level of fine-tuning in the case of \ulxielong. Regardless of the correct scenario, the detection of mHz-QPOs from both M82 X-2 and \ulxielong\ further confirms that one should avoid constraining the mass of the accretor in the ULX from the observed QPO frequency. The QPOs from both PULXs show little to no evolution in different epochs: more detections from other PULXs are needed to confirm that this is a properties of PULXs at super-Eddington luminosities.

In conclusion, we suggest that the presence of mHz-QPOs might also be a common feature of PULXs. However, the drop of the pulsed fraction when QPOs are present further complicates the detection of spin pulsations. This could also mean that the fraction of PULXs over the whole ULX population could be even higher than estimated and that the phenomenology of PULXs is more complex than previously thought. New observations targeting known as well as candidate PULXs will help us better understanding these new phenomena and testing our hypothesis.





\begin{acknowledgements}
      This study is based on observations obtained with XMM–Newton, a European Space Agency (ESA) science mission with instruments and contributions directly funded by ESA Member States and National Aeronautics and Space Administration (NASA). The scientific results reported in this article are based in part on data obtained from the Chandra Data Archive and observations made by the Chandra X-ray Observatory and published previously in cited articles.
      MI is supported by the AASS Ph.D. joint research programme between the University of Rome "Sapienza" and the University of Rome "Tor Vergata", with the collaboration of the National Institute of Astrophysics (INAF). GLI acknowledges financial support from the Italian Ministry for University and Research, through grant 2017LJ39LM (UNIAM). CS acknowledges funding from the Italian Space Agency, contract ASI/INAF n. I/004/11/4. RA and GLI acknowledge financial support from INAF through grant ``INAF-Astronomy Fellowships in Italy 2022 - (GOG)''. GLI, GARC, CP, FP, AT, AW, and PE  acknowledge support from PRIN MUR SEAWIND (2022Y2T94C) funded by NextGenerationEU and INAF Grant BLOSSOM. TPR acknowledges support from STFC as part of the consolidated grant ST/X001075/1. DJW also acknowledges support from STFC (individual grant ST/Y001060/1).
\end{acknowledgements}

%
\bibliographystyle{aa} 
\bibliography{silmarillion} 
%
\begin{appendix}
    \section{Timing analysis of \chandra\ observations}\label{appendix:Chandra}

    We followed the same steps described in Sect.~\ref{sec:LPTiming} for the timing analysis of the archival \chandra\ observations. We summarise here the main points, for the ease of the discussion. 
    
    We computed the 0.5--10\,keV PDSs with the \textsc{XRONOS} task \texttt{powspec}. We used a bin time of 3.14\,s, equal to \chandra\ time resolution, and close to the 5\,s bin time used for the \xmm\ observations. We adopted a logarithmic rebin factor of 1.20 for the three longer observations (ObsIDs 13813, 13812 and 13814). The two shorter observations (ObsIDs 15496 and 13815) needed a logarithmic rebin factor of 1.30 for a better fit due to poorer statistics. This is probably a consequence of the shorter exposure times.

    We converted the PDSs in \textsc{XSPEC} format. We found that a single Lorentzian was sufficient for a good fit of the (sub-)mHz feature in the \chandra\ PDSs. To model the whole PDSs we considered again a constant and a power-law component to account for the white and red noise, respectively. The final model used to fit the \chandra\ PDSs is described by the following equation:
    \begin{equation}\label{eq:PDSChandra}
        P(\nu)=\mathrm{const}_\mathrm{WN}+K_\mathrm{RN}\nu^{\Gamma_\mathrm{RN}}+K\frac{\Delta\nu}{2\pi}\frac{1}{(\nu-\nu_0)^2+(\Delta\nu/2)^2}
    \end{equation}
    where $P(\nu)$ is the power $P$ at the frequency $\nu$, the first two terms on the right-hand side describe the white and red noise, respectively, and the last term is the Lorentzian. $\nu_0$ is the centroid frequency and $\Delta\nu$ its full width at half maximum. The only difference from equation~\eqref{eq:PDSmodel} is the absence of the summation term, as we are considering just one Lorentzian component instead of two. 
    
    We found that the white-noise constant and the red-noise, power-law index values are consistent with the ones expected using the Leahy normalization (const$_\mathrm{WN}=2$, $-2<\Gamma_\mathrm{RN}<-1$, see \citealt{vanderKlis1989}). For observations 15496 and 13812, the fit does not require the power-law component and the continuum of the PDS can be described by the white-noise component only.

    We report the results of our fits in Tab.~\ref{tab:Chandra_PDSs}. The broad-feature parameters show little to no evolution between the \chandra\ and the \xmm\ observations (Tab.~\ref{tab:PDSsfits_LP}).
    However, while in the longer observations those parameters are more similar to those found for the broad component, in the two shorter observations (ObsIDs 15496 and 13815) the centroid frequencies are closer to those shown by the QPO at 0.5\,mHz. Nevertheless, with the available data, we could not tell if this difference is due to an evolution of the feature itself or the intrinsic resolutions of the respective PDSs.

    \begin{table*}
       \caption{Parameters of the Lorentzian obtained from the fit of the 0.5--10\,keV PDSs of the \chandra\ archival observations showing the broad feature in the mHz range. $\nu_0$: centroid frequency of the Lorentzian. $\Delta\nu$: full width at half maximum (FWHM) of the Lorentzian. $\nu_\mathrm{char}$: characteristic frequency of the Lorentzian. $Q=\nu/\Delta\nu$: approximate quality factor of the Lorentzian. rms: rms fractional variability of the Lorentzian. Errors given at 1$\sigma$ (68.3\%) confidence level. Two additional components have been added to the fit to take into account the white and red noise at high and low frequencies, respectively, unless otherwise stated.
       }
       \centering
       \begin{tabular}{ccccccc}
       \hline\hline
          ObsID & $\nu_0$ & $\Delta\nu$ & $\nu_\mathrm{char}$ & $Q$ & rms & $\chi^2$/dof\\
          & (mHz) & (mHz) & (mHz) & & (\%) & \\
        \hline
        \hline
        \\
        13813 & $0.97\apm{0.10}{0.11}$ & $1.51\apm{0.32}{0.44}$ & $1.23\apm{0.13}{0.16}$ & 0.6 & $33.1\pm2.0$ & 29.08/42 \\ \\
        13812\tablefootmark{a} & $0.94\apm{0.10}{0.10}$ & $1.34\apm{0.26}{0.33}$ & $1.15\apm{0.11}{0.13}$ & 0.7 & $39.8\pm2.1$ & 45.25/44 \\ \\
        15496\tablefootmark{ab} & $0.66\apm{0.12}{0.44}$ & $0.8\apm{0.4}{1.5}$ & $0.78\apm{0.15}{0.54}$ & 0.8 & $31.6\apm{4.2}{6.4}$ & 18.90/25\\ \\
        13814 & $0.83\pm0.11$ & $1.17\apm{0.27}{0.39}$ & $1.02\apm{0.12}{0.15}$ & 0.7 & $33.8\pm2.6$ & 25.27/42 \\ \\
        13815\tablefootmark{b} & $0.39\apm{0.12}{0.10}$ & $0.60\apm{0.20}{0.27}$ & $0.49\apm{0.12}{0.11}$ & 0.7 & $41.5\apm{4.3}{2.2}$ & 16.39/26\\ \\
        \hline
       \end{tabular}
        \tablefoot{
        \tablefoottext{a}{For this observation a two-component model (white noise plus Lorentzian in the mHz range) was sufficient to model the PDS.}
        \tablefoottext{b}{Geometric rebin changed to a factor 30\%, due to poor statistics.}
        }
       \label{tab:Chandra_PDSs}
   \end{table*}
    
\end{appendix}

\end{document}